\documentclass[twocolumn]{aastex701}

\usepackage{amsmath,amssymb,mathrsfs}
\usepackage{CJK}

\begin{document}
\begin{CJK*}{UTF8}{gbsn}

\shorttitle{BSG Envelope Simulations}
\shortauthors{Ma et al.}

\title{3D Radiation Hydrodynamics of Blue Supergiant Envelopes}

\author[orcid=0000-0001-6117-5750]{Linhao Ma (马林昊)}
\affiliation{Kavli Institute for Theoretical Physics, University of California, Santa Barbara, CA 93106, USA}
\affiliation{Department of Astrophysical Sciences, Princeton University, 4 Ivy Lane, Princeton, NJ 08544, USA}
\email{linhaoma@princeton.edu}

\author[orcid=0000-0001-8038-6836]{Lars Bildsten}
\affiliation{Kavli Institute for Theoretical Physics, University of California, Santa Barbara, CA 93106, USA}
\affiliation{Department of Physics, University of California, Santa Barbara, CA 93106, USA}
\email{bildsten@kitp.ucsb.edu}

\author[orcid=0000-0002-2624-3399]{Yan-Fei Jiang (姜燕飞)}
\affiliation{Center for Computational Astrophysics, Flatiron Institute, New York, NY 10010, USA}
\email{yjiang@flatironinstitute.org}

\begin{abstract}

We present the first 3D radiation hydrodynamical simulation of the outer envelope of a blue supergiant. These near Eddington limited massive stars in the Hertzsprung gap exhibit substantial photometric variability and have spectroscopic line widths indicating unexplained macroturbulent velocity fields. Our \texttt{Athena++} simulations find that the convection associated with the opacity peak from iron directly extends to the photosphere, as found in previous 3D simulations of massive main sequence stars. Unlike 1D stellar models, this single convecting region imprints a velocity field that is consistent with spectroscopic models of macroturbulence. In addition, the convection in the outermost layer leads to such large density variations that the exiting radiation varies on large ($\lesssim 10 \% $) amplitudes and short ($\sim$ days) timescales. This leads to stochastic low frequency (SLF) photometric variability, with amplitudes and power spectra broadly consistent with observations.

\end{abstract}


\section{Introduction} 

Blue supergiants (BSGs) are massive stars that have evolved beyond the main sequence. In single-star evolution, massive stars cross the Hertzsprung gap after leaving the main sequence, leading to relatively short predicted BSG lifetimes. This picture has difficulty explaining the observed numbers of BSGs \citep{Castro2014,Castro2018,deBurgos2023}, motivating alternative evolutionary pathways involving binaries \citep{Bellinger2023,Menon2024}.

BSG are also intriguing astronomical objects that show remarkable time variability. For many decades, it has been known that their  brightness fluctuates \citep{Bresolin2004,Moravveji2012}. Space based photometry has allowed for this variance to be well characterized as stochastic low-frequency (SLF) variability with substantial amplitudes around a few cycles per day \citep{Bowman2019,Bowman2019b,Pedersen2019,Bowman2020,Ma2024,Kourniotis2025}. Spectroscopic observations show substantial line broadening (and variability) beyond what is expected from stellar rotation, known as ``macroturbulent'' broadening \citep{Ryans2002,SimonDiaz2010,Serebriakova2023}. The physical origins of both signals remain debated and motivate our work here. 

Proposed explanations for the temporal variability of massive stars include global non-radial oscillations/waves excited in the stellar interior \citep{Lucy1976,Kaufer1997,Aerts2009,Bowman2019b}, turbulent motions associated with convection in the stellar envelope \citep{Cantiello2021}, and stellar winds \citep{Aerts2018,Ramiaramanantsoa2018,Krticka2018,Krticka2021,Bailey2024}. Distinguishing them may therefore provide new probes of the internal structure and evolutionary state of BSGs.

While the wave interpretation can be investigated using linear oscillation calculations based on one-dimensional stellar models, the convective interpretation is intrinsically multidimensional.  For hot massive stars, 1D models typically predict subsurface convection zones associated with opacity peaks, while the overlying envelope remains radiative. This has historically made surface convection a less favored explanation for the observed photometric variability and macroturbulent line broadening in massive stars (see, e.g., the introduction in \cite{Serebriakova2024} for a brief review of historical explanations).
Recent multidimensional radiation-hydrodynamic (RHD) simulations, however, have shown that turbulence originating in subsurface convection zones extends much closer to, or even reaches, the photosphere in massive main sequence stars (e.g., \citealt{Schultz2022,Schultz2023a,Debnath2024}).

In this work, we present three-dimensional RHD simulations of a BSG envelope and explore the connection of these simulations to the observations. 
We use \texttt{Athena++} to solve the nonlinear RHD equations of the convective and radiating turbulent flow from first principles. We find that convection driven near the iron opacity peak develops into large-scale turbulent motions that extend to the stellar photosphere. The result naturally produces both stochastic photometric variability and substantial velocity broadening, in qualitative agreement with the observed SLF variability and macroturbulent velocities of BSGs. These results suggest that envelope convection may provide a common physical origin for these observational signatures of blue supergiants.

This paper is organized as follows: in Section \ref{sec:initialization}, we describe the initialization of our 3D simulations, including the 1D stellar profile we used for initial conditions (\ref{sec:1d_model}) and the 3D simulation setup (\ref{sec:3d_setup}). We present the steady state solutions of our simulation in Section \ref{sec:steadystate}. We discuss the implications of the observations in Section \ref{sec:observational_implications}, including the indications of SLF variability (\ref{sec:photometry}) and the effects on macroturbulent velocities (\ref{sec:velocity}). Finally, we conclude in Section \ref{sec:conclusion}.

\section{Initializing 3D Stellar Simulations}
\label{sec:initialization}

In this section, we describe our methods for initializing three-dimensional simulations of BSG envelopes. We start from a spherically symmetric stellar structure profile obtained from 1D stellar evolutionary models (Section \ref{sec:1d_model}), and then map it into a 3D simulation domain (Section \ref{sec:domain}) with certain initial and boundary conditions (Sections \ref{sec:IC} and \ref{sec:BC}). We then use the \texttt{Athena++} code \citep{Stone2020} to numerically solve the time-dependent fluid equations with radiation transport.

\subsection{1D Stellar Profile}
\label{sec:1d_model}

\begin{figure*}
    \centering
    \includegraphics[width=\textwidth]{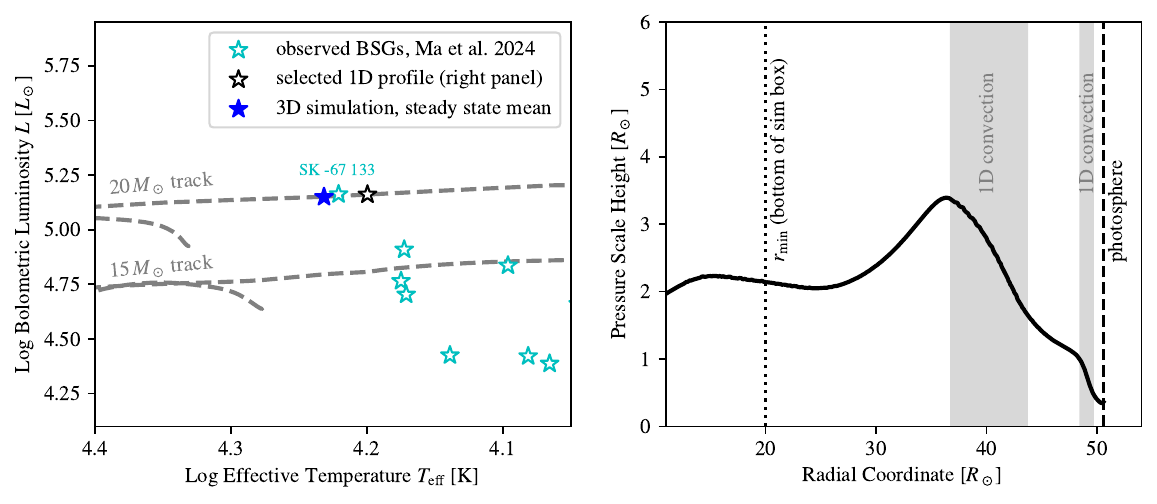}
    \caption{\textbf{Left}: the HR diagram showing observed blues supergiants (BSGs, cyan empty stars; selected from \protect\citealt{Ma2024}), two stellar tracks at different masses from 1D evolutionary models (gray dashed lines), our selected 1D profile to set up the 3D simulation (black empty star), and the mean effective temperature and luminosity from our 3D simulation after it reaches a steady state (blue filled star). The 1D profile and our 3D models closely match the BSG SK -67 133 on the HR diagram. \textbf{Right}: The stellar profile of our selected 1D model from the left panel, showing the pressure scale height (black line) and the sub-surface convective regions (gray shadows) predicted in the 1D model. The bottom of our simulation domain is set at $r_\mathrm{min}=20\,R_\odot$ (black dotted line), well below the convective regions.
    }
    \label{fig:HR}
\end{figure*}

Observationally, BSGs are identified as a distinct class by their spectral types and luminosities, living in the Hertzsprung gap of massive stars. Figure \ref{fig:HR} shows the HR diagram of 12 BSGs selected from \cite{Ma2024} in the LMC.

We start with 1D hydrostatic stellar evolutionary models for BSGs that inform our full 3D calculations. We use the MESA stellar evolution code (r24.08.1; \citealt{Paxton2011,Paxton2013,Paxton2015,Paxton2018,Paxton2019,Jermyn2023}) and run the ``BSG merger models'' from \cite{Bellinger2023}, which are relaxed from an initial chemical abundance inspired by stellar merger simulations \citep{Schneider2019}. These models are astrophysically preferred over the single massive stellar evolution models, as they help resolve the difficulties arising from the observed overabundance of blue supergiants by extending their lifetimes across the Hertzsprung gap. We find practically no difference between the envelope structures of these models and those of single star models given the same $T_\mathrm{eff}$ and $L$, as they differ only in their core chemical profiles, which do not affect their chemically homogeneous envelopes.

We use the same MESA input parameters as described in \cite{Bellinger2023} with a metallicity of $Z=0.008$, typical for stars in the LMC. We assume the star is non-rotating. The 1D models use the Rosseland mean opacities , which are calculated using the OPAL opacity tables \citep{Iglesias1993,Iglesias1996} based on the GS98 solar chemical abundances \citep{Grevesse1998}. We ran two evolutionary models with initial masses of $15\,M_\odot$ and $20\,M_\odot$ from the ignition of core helium burning to core helium depletion. Their stellar tracks on the HR diagram are shown in Figure \ref{fig:HR}. These models cover the ranges of $T_\mathrm{eff}$ and $L$ of observed BSGs. We hence select a particular stellar profile from our $20\,M_\odot$ model to establish our 3D stellar envelope model, shown as the black cross in Figure \ref{fig:HR}. This profile has $T_\mathrm{eff,MESA}=15835\,\mathrm{K}$ and $\log{L_\mathrm{MESA}/L_\odot}=5.161$. The parameters are close to one of the observed blue supergiants SK -67 133, with $T_\mathrm{eff}=16630\pm850\,\mathrm{K}$ and $\log{L/L_\odot}=5.162$.

\subsection{3D Simulation setup}
\label{sec:3d_setup}

We perform 3D radiation hydrodynamic simulations of BSG envelopes with the \texttt{Athena++} code \citep{Stone2020}. The code solves the time-dependent hydrodynamical fluid equations with an implicit scheme to couple the radiation transfer equations for specific intensities in discrete angles \citep{Jiang2021}. It is Eulerian in nature, solving the partial differential equations on pre-determined spatial grids for a fixed simulation domain with given initial and boundary conditions. 

\subsubsection{Simulation Domain and Spatial Grids}
\label{sec:domain}

As full star simulations are computationally expensive, we restrict our simulation domain to a part of the stellar envelope, which is geometrically a wedge bounded between $r_\mathrm{min}\leq r\leq r_\mathrm{max}$, $\theta_\mathrm{min}\leq\theta\leq\theta_\mathrm{max}$, and $\phi_\mathrm{min}\leq\phi\leq\phi_\mathrm{max}$ in spherical coordinates $r,\theta,\phi$. We use the 1D stellar profile, as shown in the right panel of Figure \ref{fig:HR}, to determine the coordinate ranges. The shaded regions show the two sub-surface convective layers predicted by local convective instability analysis from the 1D stellar structure, and the black line shows the pressure scale heights, which are, roughly, the size of the largest convective eddies. 

As our primary purpose is to study time-dependent convection in these regions, we choose $r_\mathrm{min}=20\,R_\odot$ (black dashed line), which is roughly $5-8$ scale heights below the lower bound of the predicted convective regions at $r\approx 36\,R_\odot$. We choose $r_\mathrm{max}=80\,M_\odot$ well above the photosphere of the 1D stellar profile ($\approx 50\,R_\odot$, black dashed line). This radial range safely covers the entire convective regions, even with the convective overshooting/undershooting of a few scale heights that are often found in 3D simulations \citep{Schultz2022,Schultz2023a}.

To make our wedge simulation representative of the whole star, we choose an angular size such that the convective motions are not expected to be spatially correlated. We chose $\pi/3\leq \theta \leq\ 2\pi/3$ and $0\leq \phi \leq\ \pi/3$ for our simulations, which cover a solid angle of $\pi/3$. If we assume the largest convective scale is roughly one scale height (maximally $\approx3\,R_\odot$ in convective regions), this angular range will cover at least $\pi/3\times40^2/3^2\sim 200$ uncorrelated convective patterns at $r=40\,R_\odot$. This allows for many upward and downward convective flows.

We construct the spatial grids with uniform spacing in $r,\theta,$ and $\phi$ coordinates. We choose a base grid that resolves at least 6 cells per local scale height in the simulation domain. Between $35\,R_\odot$ and $55\,R_\odot$, where we expect convection due to the iron opacity peak, we employ static mesh refinement to resolve $>10$ cells per scale height. This results in 196 base cells in the radial direction and 128 base cells in both the $\theta$ and $\phi$ directions. After mesh refinement, our simulation has $2560\times 16^3=10,485,760$ cells, and each cell is assigned 120 angles for the calculation of specific intensities.

\subsubsection{Initial Condition}
\label{sec:IC}

We use a 1D stellar structure profile as the initial condition (IC) for the 3D calculations, specifically, gas density, energy, and 3D velocities on each grid for hydrodynamic ICs, as well as the initial specific intensities on each grid and discrete angles for radiation ICs. However, the latter two quantities are generally not directly available from 1D hydrostatic profiles.

In 1D hydrostatic stellar evolution models, convection is treated with mixing length theory, where energy transport is described by a process that assumes a convectively unstable fluid parcel will mix with its surroundings after it adiabatically travels a parameterized mixing length in the radial direction. In addition to fixing the temperature gradient, this also estimates the convective energy flux $F_\mathrm{conv}$ from local thermodynamic properties, satisfying:
\begin{equation}
\label{eq:energy_transport}
F_\mathrm{conv}+F_\mathrm{rad}=F_\mathrm{total}=\frac{L}{4\pi r^2}\,,
\end{equation}
where $r$ is the radial coordinate, and $L$ is the total luminosity. Here, 
\begin{equation}
    F_\mathrm{rad}=-\frac{4acT^3}{3\kappa\rho}\frac{dT}{dr}\,,
\end{equation}
is the radiative energy flux under the diffusion limit, where $a$ is the radiation constant, and $c$ is the speed of light. Gas and radiation are assumed to be at the same temperature $T$, while $\rho$ and $\kappa$ are the local density and opacity of the gas.

This prescription does not provide a 3D velocity field in the convective regions. If we map the gas quantities (density and energy) from the 1D profile into the initial conditions for our 3D simulation and assume zero initial velocities, there will be no convective flux, and Equation \ref{eq:energy_transport} will not be satisfied. Hence, the initial condition is out of thermal equilibrium, and it will take a thermal time before convection is well established.

To ameliorate this issue, we take another approach by solving for a ``radiative transfer only'' profile from our 1D stellar structures, similar to the approaches in, e.g., \cite{Goldberg2022,Schultz2022}. Specifically, we take the thermodynamic quantities (density, temperature, and pressure) at a given radius inside the convectively stable regions and re-integrate the stellar structure equations outward to obtain a new spherically symmetric stellar profile, with the assumption that radiation is the only process that transports energy, i.e. $F_\mathrm{conv}=0$, and
\begin{equation}
F_\mathrm{rad,\,radiative\,only}=\frac{L_\mathrm{MESA}}{4\pi r^2}\,.
\end{equation}
This profile is identical to the original profile in convectively stable regions, as both profiles satisfy the same equations and boundary conditions. We then map this new profile into the IC for our simulations, and convection develops in the unstable regions in less than 20 days (see discussion in Section \ref{sec:relaxation}, Figure \ref{fig:relaxation}).

We calculated this radiative transfer only profile by integrating from $r=16.5\,R_\odot$ to the photosphere\footnote{The convergence to the photosphere of this integration is not always guaranteed. Practically, we adjust the inner boundary by a small amount to find numerical convergence.}, using the same equation of state and opacity tables as the original MESA model. The resulting profile is almost identical to the original profile in convectively stable regions, as expected. In convectively unstable regions of the original profile, the new profile has a different density and temperature structure, leading to a negligible increase in the envelope mass above $r_\mathrm{min}$.

We map the gas density and energy from this profile into the constructed spatial grids with radial coordinate $r_\mathrm{grid}$ between $r_\mathrm{min}$ and the photosphere radius $ r_\mathrm{photo}$, assigning them the values interpolated at $r_\mathrm{grid}$. For grid points above the photosphere, we assign them a spherically symmetric isothermal atmosphere profile, keeping their temperatures the same at $T_\mathrm{photo}$ and making their densities scale as $\rho=\rho_\mathrm{photo}\exp(-(r_\mathrm{grid}-r_\mathrm{photo})g/P_\mathrm{photo})$ until they reach the density floor value, where $g$ is the local gravity at $r_\mathrm{grid}$, and $P_\mathrm{photo}$ is the pressure at the photosphere. We set the initial velocity field to zero at every grid point.

Stellar evolution models treat radiation in the diffusion limit, with no angular-dependent specific intensities available. To construct a consistent initial specific intensity $I(\hat{n})$ in the direction $\hat{n}$ on the grids, we make use of the moment equations for specific intensities:
\begin{equation}
    \int I(\hat{n})d\Omega=cE_\mathrm{rad}=acT^4,\,
\end{equation}
\begin{equation}
    \int I(\hat{n})\,\hat{n} \cdot \hat{r}\, d\Omega=F_{\mathrm{rad},r}=\frac{L_\mathrm{MESA}}{4\pi r^2}\,,
\end{equation}
where $E_\mathrm{rad}$ is the radiation energy density, and $F_{\mathrm{rad},r}$ is the radial radiation flux. We use a ``two-zone'' initialization, assuming the specific intensities have a uniform amplitude $I_+$ if they point to the outward half-sphere ($\hat{n} \cdot \hat{r}>0$), and $I_-$ if they point to the inward half-sphere ($\hat{n} \cdot \hat{r}<0$). The above equations yield:
\begin{equation}
    I_++I_-=\frac{acT^4}{4\pi},\;I_+-I_-=\frac{F_{\mathrm{rad},r}}{\pi}\,,
\end{equation}
leading to $I_+=(acT^4/4+F_{\mathrm{rad},r})/(2\pi)$ and $I_-=(ac T^4/4-F_{\mathrm{rad},r})/(2\pi)$. For each grid point, we obtain the temperature and radial flux values by interpolating IC (with $T=T_\mathrm{photpshere}$ and $F_{\mathrm{rad},r}=0$ outside the photosphere). We then calculate $I_+$ and $I_-$ and assign them to the initial specific intensities depending on their directions. The initial radiation field constructed in this way will then be consistent with the radiation energy and flux in the spherically symmetric 1D stellar profile.

\subsubsection{Boundary Condition}
\label{sec:BC}

We set boundary conditions for our 3D simulation domain to describe the interactions of the envelope with its surroundings. As we expect the convective motions to be homogeneous on large angular scales, we set periodic boundary conditions for all variables in the $\theta$ and $\phi$ directions. This will introduce an artificial coherence to the variability caused by convection if we map our wedge simulation to the whole sphere directly, as we discuss in Section \ref{sec:photometry}.

We would like to provide an inner boundary condition at $r_\mathrm{min}$ that connects the envelope smoothly to the stellar interior. In \texttt{Athena++}, this is done by setting the gas density, gas pressure, gas velocities, and specific intensities inside layers of ``ghost cells'' that are attached outside the active simulation domain. At every integration timestep, the ghost cells are set to the desired values, such that when the full fluid equations are solved over both the active cells and the ghost cells, the solutions automatically match the boundary conditions consistently without the need for specific treatments at the boundary.

We set the gas densities and pressures in the inner boundary ghost cells by interpolating our initial profile at their radial coordinates. For specific intensities, we use similar interpolation to obtain the temperature at the ghost cells. We then use our ``two zone'' construction again with $F_{\mathrm{rad},r}=L_\mathrm{MESA}/(4\pi r^2)$, as we did for the initialization of specific intensities in the active simulation domain. This will guarantee that the inner boundary provides the same pressure support and energy transport as in the 1D stellar profile.

The remaining question is how to set the 3D velocities in the inner boundary ghost cells. As the inner boundary resides within the convectively stable regions, the cells around it should remain close to hydrostatic equilibrium with negligible velocities. Therefore, one may na\"ively set $\mathbf{v}=0$ inside all ghost cells. However, such setups do not guarantee zero velocities at the boundary because the velocities inside cells are evaluated at cell centers rather than at cell faces, which correspond to the actual boundary of the simulation domain. 

A non-zero velocity field at the boundary cell faces will cause a problem, as it will generally carry a finite mass flux across the boundary when $v_r\neq 0$. The mass will then enter the ghost cells and be eliminated in the next integration time step when the ghost cells are reset to their desired values. This will cause the envelope mass in the active simulation domain to constantly leak from the inner boundary, which is especially problematic in the initial relaxation stage of our simulation, when a transient inflow forms before convection starts to develop (see discussion in Section \ref{sec:relaxation}).

To solve this “mass leakage” problem, we impose a ``zero-mass-flux'' inner boundary condition by adjusting the radial velocities in the layer of ghost cells near the inner boundary of the simulation domain. In our simulations, hydrodynamic fluxes across cell faces are evaluated by the HLLC Riemann solver. At each step, \texttt{Athena++} first reconstructs the left and right primitive states at the inner boundary face from the neighboring cell-centered states. HLLC then uses these face states to estimate the contact-wave speed and calculate the numerical mass flux. We numerically solve for a cell-centered radial velocity $v_{r,\mathrm{ghost}}$ for each boundary-adjacent ghost cell, such that its cell-centered state, combined with that from its neighboring active cell, leads to a vanishing HLLC contact wave speed at the boundary face after the left and right primitive states are constructed. We then set $v_r=v_{r,\mathrm{ghost}}$ and $v_\theta=v_\phi=0$ in this layer of the ghost cells near the inner boundary and set $\textbf{v}=0$ in all other ghost cells. This will give zero mass flux across the inner boundary at each angular direction. 

At the outer boundary $r_\mathrm{max}$, we adopt an outflow boundary condition. The gas density, pressure, and tangential velocity components in the ghost cells are copied from the outermost layer of active cells. The radial velocity is treated similarly when $v_r>0$, while it is set to zero when $v_r<0$ to prevent gas inflow. For the radiation field, outward-propagating specific intensities ($\hat{n}\cdot\hat{r}>0$) are copied from the neighboring active cells, whereas inward-propagating intensities ($\hat{n}\cdot\hat{r}<0$) are set to zero. This prescription prevents both gas and radiation from entering the simulation domain through the outer boundary.

\subsubsection{Opacities}

We take the gray approximation in our simulations by providing frequency-averaged gas opacities to \texttt{Athena++} for radiation transport calculations. The code needs three opacities: the absorption opacity $\kappa_\mathrm{a}$, the electron scattering opacity $\kappa_\mathrm{e}$, and the Planck-mean opacity $\kappa_\mathrm{Planck}$ to couple gas with the radiation field.

For consistency with our 1D hydrostatic models, we take the same chemical abundances from our 1D stellar profile. We adopt the hydrogen and helium mass fractions of the stellar envelope, with a hydrogen mass fraction $X=0.7$ and a helium mass fraction $Y=0.292$. The abundances of metals from lithium to zinc are taken from the solar abundance pattern of \cite{Grevesse1998} and rescaled proportionally to match the total metallicity $Z=0.008$ in our 1D model. We then obtain opacity tables for this chemical mixture from the TOPS opacity database\footnote{\url{https://aphysics2.lanl.gov}}, which provides both the Rosseland-mean opacity $\kappa_\mathrm{Rosseland}$ and the Planck-mean opacity $\kappa_\mathrm{Planck}$. By interpolating these tables along the 1D stellar profile, we verify that the resulting Rosseland-mean opacities are consistent with those used in MESA.

Since the photon energies in our stellar envelope simulations are well within the Thomson limit, we adopt the Thomson cross section for electron-scattering opacity, which gives $\kappa_\mathrm{e}=0.2\,(1+X)\,\mathrm{cm}^2\mathrm{g}^{-1}=0.34\,\mathrm{cm}^2\mathrm{g}^{-1}$. The absorption opacity can then be calculated by subtracting the electron scattering opacity from the Rosseland-mean opacity:
\begin{equation}
\kappa_\mathrm{a}=\kappa_\mathrm{Rosseland}-\kappa_\mathrm{e}.
\end{equation}
If $\kappa_\mathrm{Rosseland}<(0.34+10^{-5})\,\mathrm{cm}^2\mathrm{g}^{-1}$, we take $\kappa_\mathrm{e}=\kappa_\mathrm{Rosseland}-10^{-5}\,\mathrm{cm}^2\mathrm{g}^{-1}$ to ensure positive opacities everywhere. We then provide the opacity tables for $\kappa_\mathrm{a}$, $\kappa_\mathrm{e}$, and $\kappa_\mathrm{Planck}$ to \texttt{Athena++}, which interpolates gas density and temperature to obtain the opacities required for the calculations of radiation transport.

\subsubsection{Additional Simulation Setups}

We use an ideal gas equation of state with a mean molecular weight $\mu=0.62$, appropriate for the fully ionized material in the envelope. We assume an adiabatic index of $\gamma=5/3$ for the gas.

The full envelope outside $20\,R_\odot$ has a mass less than $0.2\,M_\odot$ in our stellar models, which is less than 1\% of the total mass of the star. Therefore, we ignore the self gravity of the envelope and assume a steady gravitational potential from a point source with $M=20\,M_\odot$ at $r=0$. This is done by setting the ``\texttt{GM}'' parameter in \texttt{Athena++} and drastically reduces the computational cost for gravity solvers.

For numerical stability, we set the Courant-Friedrichs-Lewy (CFL) number to be 0.3 for hydrodynamical calculations and 0.025 for radiation transport calculations. With this setup, our simulation generally takes less than 10 iterations to reach our preset error tolerance of $10^{-6}$ each step, ensuring good numerical convergence.

\section{Steady State Solutions}
\label{sec:steadystate}

In this section, we show the results of our 3D simulations. \ref{sec:relaxation} discusses the initial relaxation stage and the criterion for determining the steady state. \ref{sec:convection} shows the convective motions for a representative snapshot in the steady state.

\subsection{Initial Relaxation Stage}
\label{sec:relaxation}

\begin{figure*}
    \centering
    \includegraphics[width=\textwidth]{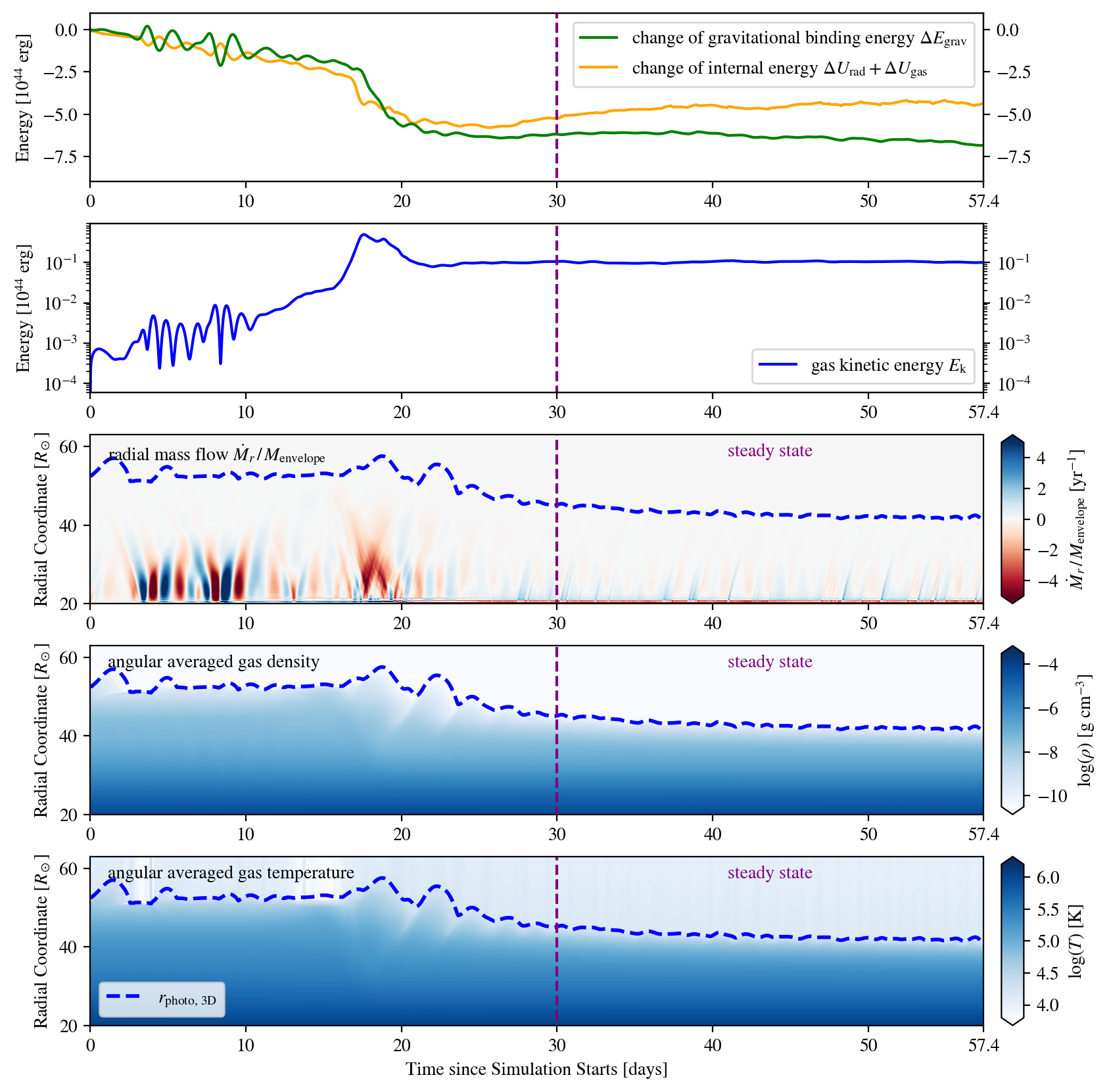}
    \caption{Several quantities varying over time in the initial relaxation stage in our 3D simulation. The purple dashed line shows $t_\mathrm{steady}=30\,\mathrm{days}$, after which we the simulation reaches a steady state. \textbf{Top two panels}: The change of total gravitational binding energy, total internal energy (for both gas and radiation) and total gas kinetic energy in our simulation domain. In the initial relaxation stage, the excess of gravitational and thermal energy are released as the stellar envelope leaves the unstable equilibrium. The energies are either transferred to the gas kinetic energy, or radiate away from our simulation domain. We see the thermal and gravitational binding energies indeed drop until $t\approx 25\,\mathrm{days}$, while the kinetic energy increases. \textbf{Third panel}: the shell-summed total radial mass flow $\dot{M}_r$ at each radius, normalized to the total envelope mass $M_\mathrm{envelope}$. In the initial relaxation stage, a huge transient radial mass flow is created in the envelope, and bounced back at the inner boundary at $r=20\,R_\odot$ as we enforce the ``zero-mass-flux'' boundary conditions. This transient flow largely damps out after $t_\mathrm{steady}$. \textbf{Bottom two panels}: The angular averaged gas density and temperature at each radius, which reach a steady state after $t_\mathrm{steady}$. The dashed blue line shows the 3D photosphere $r_\mathrm{photo,\,3D}$ by taking $\bar{\tau}(r_\mathrm{photo,\,3D})=1$, where $\bar{\tau}$ is the mean optical depth averaged over all angular directions.
    }
    \label{fig:relaxation}
\end{figure*}

Our 3D simulation starts with the convectively unstable initial condition discussed in Section \ref{sec:IC}. In the top two panels of Figure \ref{fig:relaxation}, we show the evolution of the total gravitational binding energy, radiation and gas internal energy, and the gas kinetic energy since the simulation started at $t=0$. We show the first two energies as their differences compared to the initial profile, while the initial kinetic energy is zero. As the initial model is convectively unstable, we expect a new configuration to emerge with a different energy.

We see that the gravitational binding energy of the fluid, as well as the total internal energy, indeed drops by $\approx 5\times10^{44}\,\mathrm{erg}$ in the first 20 days of evolution. This is roughly 1\% of the initial gravitational binding energy ($\approx 5\times10^{46}\,\mathrm{erg}$), or 5\% of the initial internal energy ($\approx 10^{46}\,\mathrm{erg}$) in the initial profile. A small part of this energy is transferred into the kinetic energy of the gas, which reaches $\approx 10^{43}\,\mathrm{erg}$ at $t=20\,\mathrm{days}$. The remaining majority of the energy is radiated away from the outer boundary as excess luminosity. All energies reach a roughly steady value after $t\sim 25\,\mathrm{days}$.

To understand the fluid motions in the initial relaxation stage, we show the total radial mass flow $\dot{M}_r(r)=\int r^2\rho  v_r d\Omega$ in the third panel of Figure \ref{fig:relaxation}, normalized by the total envelope mass $M_\mathrm{envelope}$. In the first 10 days after the simulation starts, we see huge radial flows forming below $r\approx 40\,R_\odot$ and often exceeding $\dot{M}_r/M_\mathrm{envelope}\simeq 4\,\mathrm{yr}^{-1}$. The flow corresponds to the initial transient as the excess energy is released, and it bounces back at the inner boundary where we enforce the ``no-mass-flux'' boundary condition.

As time evolves, the initial transient flow dampens as the kinetic energy is transferred, corresponding to the oscillating $E_\mathrm{k}$ in the second panel of Figure \ref{fig:relaxation}. The energy becomes either internal energy through gas compression, shock heating, radiation transport, and numerical dissipation, or gravitational binding energy through the displacement of the fluids. After the final huge inflow at $t\approx 18\,\mathrm{days}$, the remaining radial mass flow may correspond to the waves excited by convective motion.

As the simulation evolves into a steady state, we generally expect the angular averaged stellar profiles to be static. In the bottom two panels of Figure \ref{fig:relaxation}, we show the angular averaged gas density and gas temperature at each radius and time, which are defined in Appendix \ref{app:radial_profiles}. We also show the angular averaged photosphere $r_\mathrm{photo,\,3D}$ in the bottom three panels, defined in \ref{app:photosphere}. We can see that after the energies become static after $t=20\,\mathrm{days}$, there are still some transient oscillations near the stellar photosphere that make it unstable. The oscillations only seem to damp out after $t>30\,\mathrm{days}$ when the photosphere becomes stable.

Therefore, we set $t_\mathrm{steady}=30\,\mathrm{days}$ (purple dashed line) as the starting point of the steady state profile, upon which we conduct our analysis of the simulation. For our photometric analysis in Section \ref{sec:photometry}, we run the simulation further until $t=57.4\,\mathrm{days}$ to include one TESS sector (27.4 days) in the steady state. Despite some slight shrinking of the envelope as the bottom of our simulation domain is reaching thermal equilibrium, we verified that our analysis in Section \ref{sec:convection} and \ref{sec:velocity} does not depend on the specific snapshot we chose.

\subsection{Surface Convection}
\label{sec:convection}

\begin{figure*}
    \centering
    \includegraphics[width=\textwidth]{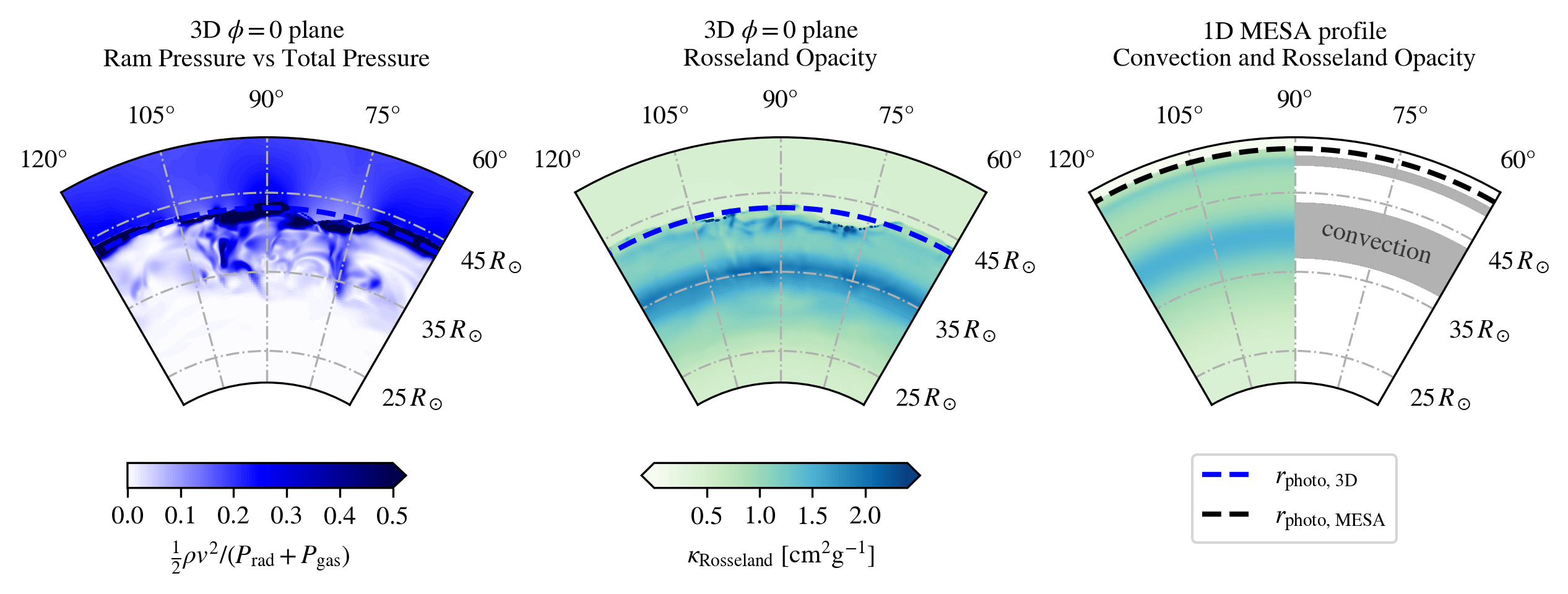}
    \caption{ \textbf{Left and Middle}: The $\phi=0$ plane ram pressure (compared to total gas and radiation pressure) and the Rosseland mean opacities for a steady-state snapshot at $t=40\;\mathrm{days}$ in our 3D simulation. We can see a significant ram pressure above $\approx 33\,R_\odot$, indicating strong convective motions. The convective region corresponds to the iron and helium opacity peaks at $\approx 35\,R_\odot$ and $\approx 42\,R_\odot$, and it extends all the way to the stellar photosphere at $\approx43\,R_\odot$. \textbf{Right}: The 1D spherically symmetric MESA profile we take (black empty star in Figure \ref{fig:HR}) projected on to the $\phi=0$ plane in our simulation domain for comparison. The gray regions show two distinct sub-surface convective regions corresponding to the opacity peaks which is different from the 3D simulation. The 1D photosphere also extends to further radius at approximately $51\,R_\odot$.
    }
    \label{fig:slice}
\end{figure*}

\begin{figure*}
    \centering
    \includegraphics[width=\textwidth]{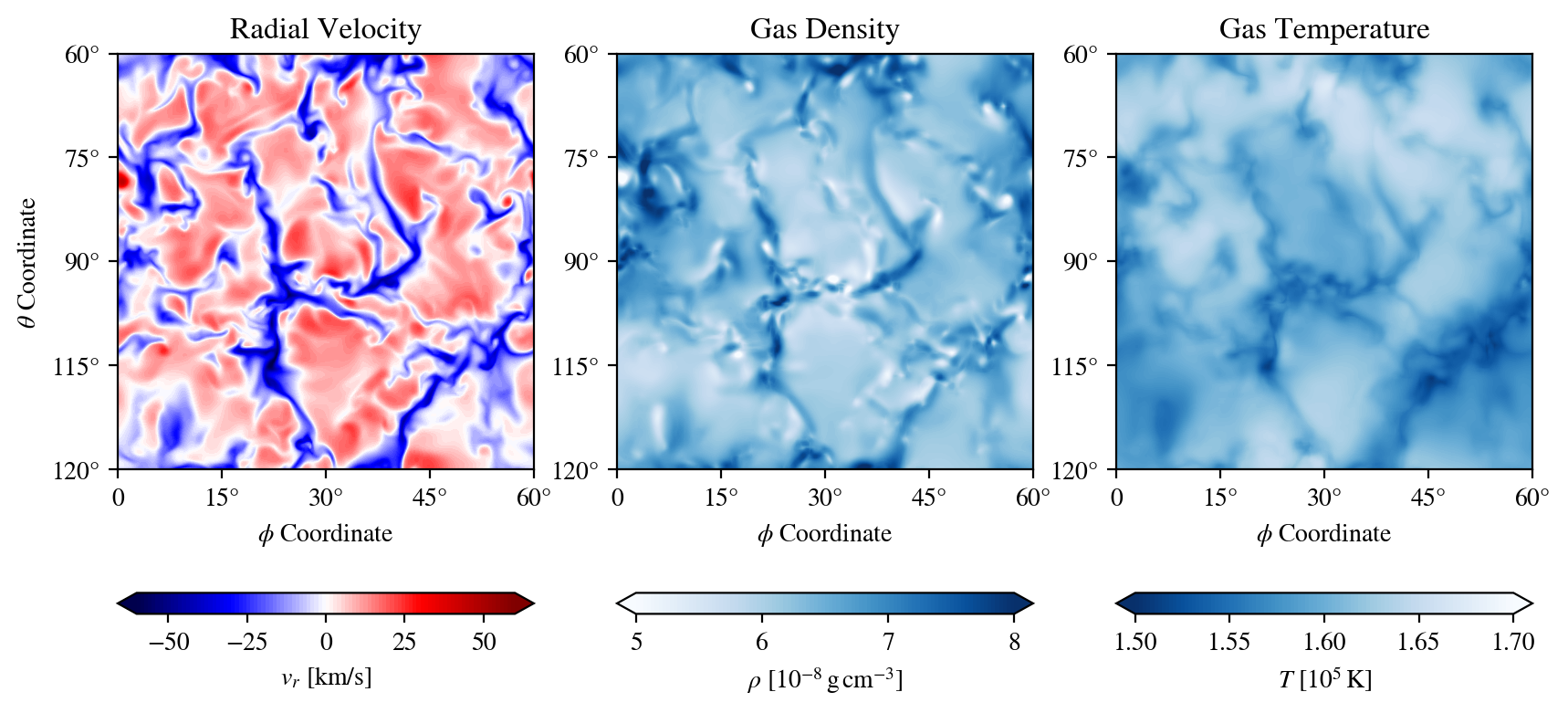}
    \caption{The $\theta-\phi$ plane at $r=36\,R_\odot$ for a steady state snapshot at $t=40\,\mathrm{days}$, showing the radial velocity $v_r$ (\textbf{left}), gas density $\rho$ (\textbf{middle}), and gas temperature $T$ (\textbf{right}). We see distinctive  convective  motions, with hot parcels moving up ($v_r>0$, lighter regions in the right two panels) and the colder parcels sinking ($v_r<0$, darker regions in the right two panels). Radiative transport blurs out the small-scale features in gas temperature.
    }
    \label{fig:vr}
\end{figure*}

To illustrate the convective motions, we take a snapshot at $t=40\,\mathrm{days}$, which is a representative steady state solution. Figure \ref{fig:slice} shows the 2D profiles evaluated at the $\phi=0$ plane in this snapshot. The left panel shows the ratio between the ram pressure of the fluid ($P_\mathrm{ram}=\frac12\rho v^2$) and the total gas and radiation pressure, which indicates the strength of hydrodynamical convective flows compared to the background pressure support. 

The ram pressure starts to become a significant fraction of the total pressure beyond $r\approx 33\,R_\odot$, indicating strong convective motions. In the middle panel of Figure \ref{fig:slice}, we show the $\phi=0$ plane Rosseland mean opacities at each cell, exhibiting a strong peak at $r\approx 35\,\mathrm{R_\odot}$. This opacity peak is caused by the partial ionization of iron at around $200,000\,\mathrm{K}$. Convection is expected to develop around this peak as it drives convective instability, exactly as we find in our simulations.

Figure \ref{fig:vr} shows the radial velocity, gas density, and gas temperatures on a 2D plane in the $\theta$ and $\phi$ directions at $r=36\,R_\odot$ for the same snapshot, just above the iron opacity peak.   The radial velocity panel (left) shows a rich spectrum of upward and downward convective motions, indicating that our simulation domain contains many convective parcels. The downward regions correspond to colder and denser (heavier) gas, while the upward regions correspond to hotter and less dense (lighter) gas, as expected from convection. In addition, the small-scale structures in the temperature profile are blurred out due to efficient radiative diffusion.

\begin{figure}
    \centering
    \includegraphics[width=\columnwidth]{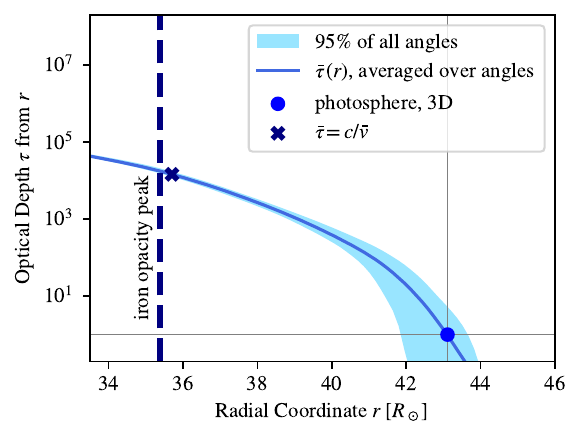}
    \caption{The optical depth $\tau(r)$ from the outer boundary in our simulation box to given radial coordinates along fixed angular directions. The shaded region shows the range of the median 95\% for the steady-state snapshot at $t=40\,\mathrm{days}$, and the solid line shows the the angular averaged optical depth $\bar{\tau}(r)$ (see definition in Appendix \ref{app:photosphere}), which yields a 3D photosphere radius of $43\,R_\odot$ at $\bar{\tau}=1$ (blue dot). However, the local photosphere at each angular direction varies from $42$ to $44\,R_\odot$ due to the turbulent motions. The cross shows where $\bar{\tau}=c/\bar{v}$ above which energy transport by convection becomes inefficient, which is just above the iron opacity peak at $\approx 35\,R_\odot$ (vertical dashed line).}
    \label{fig:tau}
\end{figure}

In the radiation pressure dominated regime of BSG envelopes, the efficiency of convective flows to transport heat is determined by the ratio of the local optical depth to $c/\bar{v}$, where $\bar{v}$ is the angular-averaged velocity amplitude of the flow (defined in Appendix \ref{app:radial_profiles}). Above this optical depth, heat transport is dominated by radiative diffusion. We find that this critical depth is just above the iron opacity peak in our simulation (Figure \ref{fig:tau}), indicating that exiting radiation diffuses through the turbulent density profile of the outermost BSG envelopes. 

To compare with 1D hydrostatic models, we show the 1D (spherically symmetric) MESA profile of the BSG envelope (black filled star in Figure \ref{fig:HR}) projected onto our simulation domain in the right panel of Figure \ref{fig:slice}. The iron opacity peak resides at a slightly larger radius ($r\approx 40\,R_\odot$) compared to our 3D simulation, which produces a sub-surface convective layer between $35\,R_\odot$ and $45\,R_\odot$. Another thin sub-surface convective region exists just below the 1D photosphere (black dashed line), corresponding to another opacity peak caused by helium partial ionization at  roughly $50,000\,\mathrm{K}$. This opacity peak is also seen in our 3D simulation around $43\,R_\odot$.

We find, however, that the convective regions in our 3D simulation are different from those in the 1D profile. Specifically, we do not see two distinct sub-surface convective regions corresponding to the two opacity peaks as predicted by 1D models. Instead, a large convective region develops from the iron peak and reaches all the way to the stellar surface. This is because the real convective flows with large inertia can still have significant momentum in the convectively stable layers predicted by the 1D model, which causes the two convective zones to merge and reach the photosphere.

\cite{Schultz2022} finds that this ``merging subsurface convection'' phenomenon generally exists in the envelopes of more massive, hotter, and less evolved stars. In those stars, the outer $\approx 1\,R_\odot$ of the envelope can be fully convective. Our simulation extends their studies to cooler objects, and the surface convective regions we find can be as deep as $\approx 10\,R_\odot$.

This phenomenon affects the morphology of the BSG photosphere. Figure \ref{fig:tau} shows the range of radial optical depth integrated along each angular direction in our simulations, and the solid line shows the angular averaged optical depth (see definitions in Appendix \ref{app:photosphere}). While the mean optical depth indicates a photosphere radius of $\approx 43\,R_\odot$, the photosphere along each individual direction ranges from $42$ to $44\,R_\odot$ due to the turbulent convective motions that rise and fall on scales of a few $R_\odot$ near the photosphere, as seen in Figure \ref{fig:slice}.

This amplitude of photospheric turbulent flow has also been found in 3D radiation-hydrodynamical simulations of red supergiants \citep{Goldberg2022,Ma2025}. In those cooler stars, surface convection occurs on even larger scales, and can sometimes make the star significantly non-spherical. Our simulation shows that this happens in BSGs as well, though the deformation cause by surface convection varies the stellar radius by $\approx 5\%.$

\section{Observational Implications}
\label{sec:observational_implications}

We now discuss two important observational implications related to the turbulent photospheres in our BSG simulations: the stochastic low-frequency variabilities (SLFs) in photometry (\ref{sec:photometry}), and the macroturbulent velocities found in spectroscopy (\ref{sec:velocity}).

\subsection{SLF Photometric Variability}
\label{sec:photometry}

\begin{figure*}
    \centering
    \includegraphics[width=\textwidth]{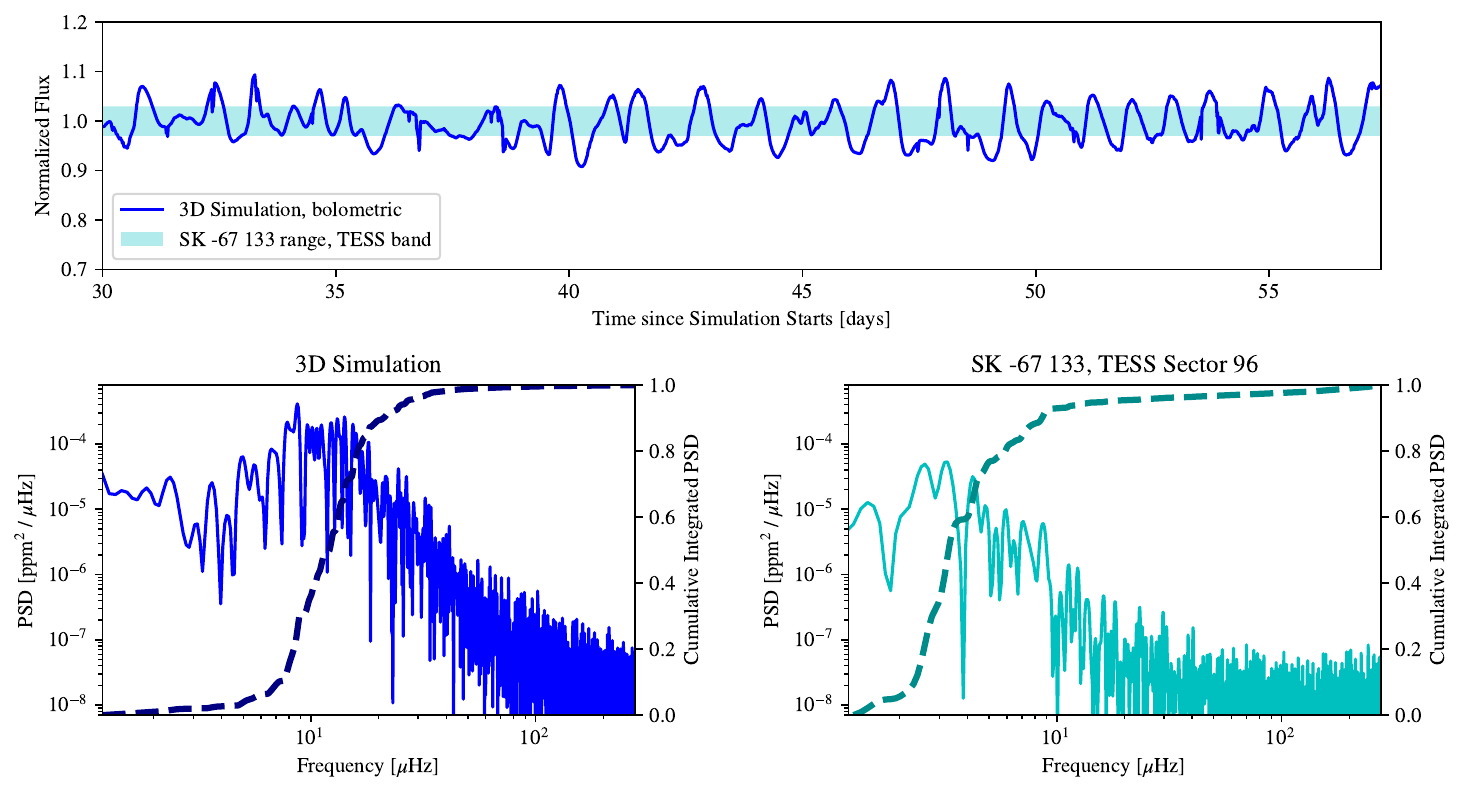}
    \caption{\textbf{Top panel}: The synthetic light curve (normalized flux) calculated from the bolometric luminosity obtained from the steady-state 3D simulation, showing up to 10\% stochastic variability in a 27.4-day observing sector (one TESS sector). The cyan region shows the typical range of TESS band variability for BSG SK -67 133, which is smaller than the simulation as expected for TESS band reductions for BSGs. \textbf{Lower left panel}: the power spectral density obtained from the synthetic light curve from the simulation, with the total integrated power as a function of frequency on the right axis. We see typical red noise signals with a characteristic peak frequency at approximately $10-20\,\mathrm{\mu Hz}$. \textbf{Lower right panel}: the power spectral density as observed by TESS sector 96 for SK -67 133, which shows similar stochastic low frequency variabilities as from the simulation, though the characteristic peak frequency seems lower, at approximately $2-4\,\mathrm{\mu Hz}$.
    }
    \label{fig:lightcurve}
\end{figure*}

Over the past decades, space-based missions such as TESS \citep{TESS} have enabled long-term photometric monitoring of BSGs. A ubiquitous signal known as stochastic low-frequency (SLF) variability has been identified \citep{Bowman2019,Bowman2019b,Bowman2020,Ma2024,Kourniotis2025}. This signal appears as an excess of power at frequencies below a few cycles per day. Several origins have been proposed for SLF variability, including internal gravity waves excited in the stellar interior \citep{Shiode2013,Rogers2013,Bowman2019}, turbulent motions associated with surface convection \citep{Cantiello2021,Schultz2022}, and the rotational modulation of structured stellar winds \citep{Aerts2018,Ramiaramanantsoa2018,Krticka2018,Krticka2021,Bailey2024}.

Our BSG simulation allows us to probe the second scenario above. Specifically, our simulation shows that sub-surface convection caused by the iron opacity peak reaches the stellar photosphere, a feature that is not present in 1D modeling (Figure \ref{fig:slice}). The radiation must then diffuse through the time dependent, inhomogeneous medium created by convective turbulence. 

To compare with the observations, we now generate a synthetic light curve for a baseline of 27.4 days (i.e., one TESS sector) from the steady-state realm of our simulation. We first calculate the bolometric luminosity $L_\mathrm{bol}(t)$  by integrating the total radial radiation flux in the outermost layer of the simulation domain (see details in Appendix \ref{app:luminosity}) for a sequence of snapshots separated by a cadence of $\Delta t = 0.0134$ days. This allows us to probe the power spectrum up to a Nyquist frequency of $1/(2\Delta t)=430\,\mu\mathrm{Hz}$.

We first compare the amplitude of the photometric variability to the observed signal. To obtain a normalized flux, we perform a linear fit to the bolometric luminosity over time to get a mean luminosity $L_\mathrm{bol,\,mean}(t)$. This allows us to remove a slight decreasing component in the mean flux as the bottom of the BSG envelope reaches thermal equilibrium. The normalized flux is then calculated from
\begin{equation}
    f_\mathrm{norm}(t)\equiv \frac{\Delta F}{F}=1+\sqrt{\frac{\Omega}{4\pi}}\left(\frac{L_\mathrm{bol}}{L_\mathrm{bol,\,mean}}-1\right)\,.
\end{equation}
Here, $\Omega=\pi/3$ is the total solid angle of the simulation domain, and the factor $\sqrt{\Omega/4\pi}$ is the amplitude reduction we apply to account for the decoherence expected from a full star \citep{Schultz2022}. In simpler terms, if all local patches are uncorrelated, the local variability above a small patch of the star will be larger than what one would see from the entire star. 

The top panel of Figure \ref{fig:lightcurve} shows the synthetic normalized flux, exhibiting a rich spectrum of stochastic variabilities with no apparent periodicity. The amplitude after coherency reduction can reach up to 10\% of the mean bolometric flux. This is higher than the typical range for BSGs from TESS observations (up to 3\% to 4\%), and reflects that the 
TESS band\footnote{$\lambda=600-1000\,\mathrm{nm}$} only observes the Rayleigh-Jeans tail of the full black body spectrum for hot stars like BSGs. 

Specifically, the power observed in the TESS band can be integrated to yield a linear dependence of the total flux on temperature:
\begin{equation}
    F_\mathrm{TESS}\propto 2ck_\mathrm{B}T\int_\mathrm{TESS\,band}\lambda^{-4}d\lambda,
\end{equation}
while the bolometric flux satisfies the Stefan-Boltzmann Law, which is quartic in temperature:
    $F_\mathrm{bol}\propto\sigma_\mathrm{SB} T^4$. Therefore, a small temperature perturbation $\Delta T$ in the photosphere produces bolometric flux variability with amplitudes of $\Delta F_\mathrm{bol}/F_\mathrm{bol}=4\Delta T$, while in the TESS band, the flux amplitudes are $\Delta F_\mathrm{TESS}/F_\mathrm{TESS}=\Delta T$. This reduction of a factor of a few for the TESS band places our simulated variability well in the realm of what's observed. 

Next, we compare the power spectral density (PSD) of the synthetic light curve from the simulation to the data. The lower panels of Figure \ref{fig:lightcurve} show the Lomb-Scargle periodogram of the synthetic light curve and the observed flux from TESS sector 96 for BSG SK -67 133, which is the closest BSG match to the 3D simulation on the HR diagram (see Figure \ref{fig:HR}). We used the Python package Lightkurve \citep{Lightkurve} to download the TESS light curves for this star from the Mikulski Archive for Space Telescopes (MAST\footnote{https://archive.stsci.edu/}), using the data processed by a pipeline developed by the Science Processing Operations Center (SPOC; \citealt{Jenkins2016}). We obtained both periodograms using the \texttt{timeseries.LombScargle} function in the \texttt{astropy} Python package \citep{astropy:2013,astropy:2018,astropy:2022}. The right axis of both panels shows the corresponding integrated fractional power from $1.25\,\mu\mathrm{Hz}$ to $278\,\mu\mathrm{Hz}$ following the approach of \cite{Pedersen2025}. 

The synthetic light curve has a red noise PSD similar to the SLF observed for SK -67 133. In both periodograms, we observe an excess of power at low frequencies, with an exponentially decaying tail as the frequency increases. However, the characteristic peak frequency of the simulated power differs from that of the observation, as seen in the fractional integrated powers. Approximately 80\% of the observed red noise power is below $6-7\,\mu\mathrm{Hz}$, while there is still significant power above this frequency in the synthetic light curve, making the 80\%-power threshold frequency roughly $20\,\mu\mathrm{Hz}$, a factor of 3 higher. 

We do not know the origins of this discrepancy. It can be caused by the stochastic nature of this signal, as the particular sector from the observation/simulation may not be the most representative. To check this, we analyzed the characteristic frequencies of all 37 available TESS sectors from SK -67 133, and we find that the characteristic peak frequency can indeed vary by a factor of a few (though none of them exceed $10\,\mu\mathrm{Hz}$). Our simulation may also have limitations due to finite resolutions. Nevertheless, the simulation shows that surface convection can qualitatively produce the SLF signal observed in BSGs.

We note that our simulations cannot rule out an internal-gravity-wave origin for the SLF signal alone (e.g., \citealt{Bowman2019}). If such waves are excited in the convective core and propagate through the deep stellar interior, most of their propagation cavity lies outside our envelope-focused simulation domain, making them unable to be captured in our models. \cite{Anders2023} showed that such waves, although usually producing surface variability amplitudes below current detection limits, can in principle generate a similar red-noise spectrum. However, our simulations disfavor another proposed scenario in which internal gravity waves are trapped between the iron- and helium-opacity-peak convection zones \citep{Ma2024}. We find that these convection zones connect with each other and extend to the photosphere, leaving no intervening propagation cavity in which gravity waves could be trapped.

\subsection{Macroturbulent Velocities}
\label{sec:velocity}

\begin{figure}
    \centering
    \includegraphics[width=\columnwidth]{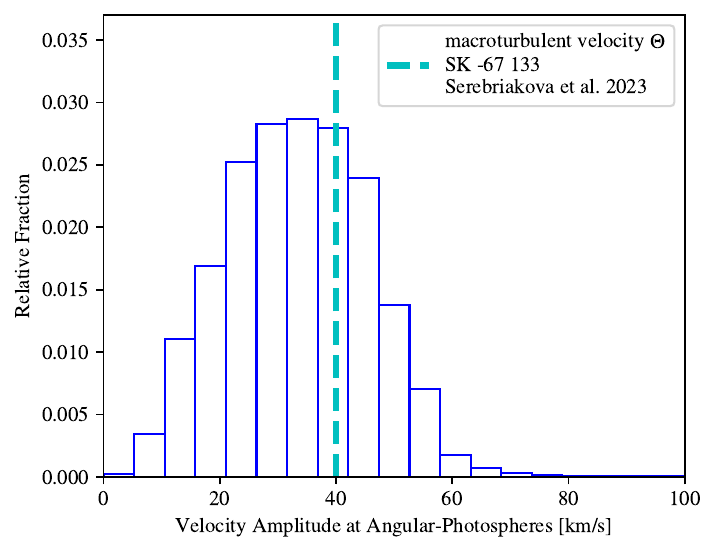}
    \caption{A histogram of the velocity amplitude obtained at the local angular-photosphere $r_\mathrm{photo}(\theta,\phi)$ at each direction, from a snapshot at $t=40\,\mathrm{days}$. The dashed line shows the macroturbulent velocity $\Theta$ measured for BSG SK -67 133. The velocities distribute around a median of $30-40\,\mathrm{km/s}$, potentially explaining  the macroturbulent velocities inferred for BSGs.}
    \label{fig:velocity}
\end{figure}

Spectroscopic observations of BSGs have long revealed substantial line broadening in excess of what can be caused by stellar rotation \citep{Howarth1997,Ryans2002,SimonDiaz2010,Serebriakova2023}. This additional broadening is commonly referred to as ``macroturbulence,'' characterized by velocities that can reach $\sim10-100\,{\rm km\,s^{-1}}$. The physical origin remains uncertain, with proposed explanations including the collective effects of non-radial pulsations and turbulent motions associated with subsurface convection \citep{Aerts2009,Grassitelli2015,Serebriakova2024}.
As convective motion reaches the photosphere in our steady-state solutions (Section \ref{sec:convection}), the latter interpretation can be tested.

Figure \ref{fig:velocity} shows the histogram of the photosphere velocity amplitudes in a $t=40\,\mathrm{days}$ steady-state snapshot from our simulations. The velocities are evaluated at the local photosphere $r_\mathrm{photo}(\theta,\phi)$ in each angular direction $\theta$ and $\phi$, such that the optical depth at $r_\mathrm{photo}(\theta,\phi)$ along the radial direction is 1. As the steady state implies a zero mean velocity, the velocity amplitudes can be understood as the rms velocities that contribute to line broadening. The velocities have a broad distribution between $\approx 10-70\,\mathrm{km}/\mathrm{s}$, with a median around $30-40\,\mathrm{km}/\mathrm{s}$. 

\cite{Serebriakova2023} report the macroturbulent velocities $\Theta$ for 126 BSGs in the LMC based on spectroscopic analysis. They find that $\Theta$ typically lies between $10-100\,\mathrm{km}/\mathrm{s}$, consistent with the photosphere velocity distribution in our simulation. Specifically, they measured a $\Theta\approx40\,\mathrm{km/s}$ for the BSG SK -67 133, whose stellar parameters match closest with our simulation (see Figure \ref{fig:HR}). The value closely matches the median of photosphere velocities in the steady state snapshot shown in Figure \ref{fig:velocity}, making surface convection originating from the iron opacity peak  a very promising explanation for the measured macroturbulent line broadening in BSGs.

While it is usually assumed that BSGs with radiative envelopes allow for a clean interpretation of $v\sin i$ as rotational velocities, our simulation shows that this is likely not the case. The convective velocities can cause additional line broadening of up to $100\,\mathrm{km/s}$, surpassing the effects of rotational broadening.

\section{Conclusion and Future Work}
\label{sec:conclusion}

In this work, we achieved the first 3D RHD simulation of a blue supergiant (BSG) envelope. The simulation solved the full 3D hydrodynamics and radiation transfer equations in a wedge region that covers a portion of the envelope, with no free parameters. This simulation reveals the turbulent dynamics in the surface convective zones of BSGs, which are excited by the local convective instability near the iron opacity peak.

We show that the convective motions in our simulation extend all the way to the photosphere, allowing convective plumes to reach the radiating surface. This affects the morphology of the photosphere. Specifically, the photosphere radius can vary by $2\,R_\odot$, or 5\% of the total stellar radius in our simulation. This makes the BSG non spherically symmetric, in a manner similar to (but less extreme than) the findings from simulations of red supergiants \citep{Goldberg2022,Ma2025}.

The radiation flux in the BSG envelope travels through the inhomogeneous medium caused by surface convection, creating stochastic variabilities in the BSG luminosity. We calculate a synthetic light curve for a baseline of 27.4 days (one TESS sector) and compare it with the observed TESS photometric flux for BSG SK -67 133, which closely matches our simulated stellar parameters in the HR diagram (Figure \ref{fig:HR}). The synthetic light curve shows stochastic low frequency (SLF) variability similar to what is found in data, in qualitative agreement with the variability amplitude and the power spectrum. Our simulation hence shows that surface convection can be the origin of SLF signals found in BSGs.

The convective flows reaching the BSG surface imprint a velocity field at the stellar photosphere, which can provide excess ``macroturbulent'' line broadening in addition to rotational broadening. We found that the velocities at local angular-photospheres in our simulation have a broad distribution between $\approx10-70\,\mathrm{km/s}$ with a median near $\approx 40\,\mathrm{km/s}$. This is in remarkable agreement with the macroturbulent velocities found in spectroscopic observations for BSG SK --67 133 ($\approx 40\,\mathrm{km/s}$, \citealt{Serebriakova2023}), and for other BSGs with typical values between $10-100\,\mathrm{km/s}$. Surface convection can hence explain the macroturbulent line broadening for BSGs.

There are improvements that can be made to our work. We use static mesh refinement in the convective regions, which cannot always guarantee enough resolution near the photosphere. This limits the ability of our simulation to achieve a time dependent multi-color light curve or spectra. Future multi-group RHD simulations with higher spatial resolution near the photosphere would achieve these goals.

\begin{acknowledgments}
We thank Eliot Quataert and Selma de Mink for early discussions as this work was being defined. We thank Jared Goldberg, Samuel Boos, and Yixian Chen for useful discussions. This research was supported in part by grant NSF PHY-2309135 to the Kavli Institute for Theoretical Physics (KITP) and by grant ATP-80NSSC22K0725 through the NASA Astrophysics Theory Program. 
LM is supported by the Gordon and Betty Moore Foundation Grant GBMF7392 at the Kavli Institute for Theoretical Physics (KITP) and the Lyman Spitzer, Jr. Postdoctoral Fellowship of Princeton University. This work also benefited from the Mitchel Postdoctoral Scholar Career Development Fund at KITP and interactions with a variety of researchers that were funded by the Gordon
and Betty Moore Foundation through grant GBMF5076.01. Computational resources were provided by
the Flatiron Institute.

\end{acknowledgments}

\facilities{Some of the data presented in this paper were obtained from the Mikulski Archive for Space Telescopes (MAST) at the Space Telescope Science Institute. The specific observations analyzed can be accessed via \dataset[https://doi.org/10.17909/df38-ax53]{https://doi.org/10.17909/df38-ax53}. STScI is operated by the Association of Universities for Research in Astronomy, Inc., under NASA contract NAS5–26555. Support to MAST for these data is provided by the NASA Office of Space Science via grant NAG5–7584 and by other grants and contracts.}

\software{This research made use of the following Python packages: Astropy \citep{astropy:2013, astropy:2018, astropy:2022}, Lightkurve, \citep{Lightkurve}, numpy \citep{Harris2020}, and SciPy \citep{scipy}. The stellar models make use of \texttt{MESA}, r24.08.1 \citep{Paxton2011,Paxton2013,Paxton2015,Paxton2018,Paxton2019,Jermyn2023}, and \texttt{Athena++} \citep{Stone2020}.}

\appendix

\section{Definitions of Global and Mean Properties}
\label{app:mean_define}

\texttt{Athena++} solved the time-dependent RHD equations on a fixed 3D spatial grid in spherical coordinates, with  $N_r,\,N_\theta,$ and $N_\phi$ cells in each direction. For each cell in the simulation domain denoted by indices $1\leq i \leq N_r$, $1\leq j \leq N_\theta$, and $1\leq k \leq N_\phi$ (where indices increase with coordinates), the solution provides the volume-centered value $Q_{ijk}$ for a given primitive variable $Q$ at the cell centered coordinate $(r_i,\theta_j,\phi_k)$. We now summarize the definitions of the global and mean properties of the simulated stellar envelope.

\subsection{Total Energies}
\label{app:energy}

The total energies are calculated by summing over all energies in the simulation box. The total kinetic energy of the gas is given by:
\begin{equation}
    E_\mathrm{k}=\sum_{ijk}\frac{1}{2}\rho_{ijk} v_{ijk}^2\Delta V_{ijk}\,,
\end{equation}
where $\rho$ is the gas density, $v^2=v_r^2+v_\theta^2+v_\phi^2$ is the square of the total gas velocity. $\Delta V$ is the volume of the cell, given by:
\begin{equation}
    \Delta V_{ijk}\equiv\int_{\phi_{ijk,-}}^{\phi_{ijk,+}}\int_{\theta_{ijk,-}}^{\theta_{ijk,+}}\int_{r_{ijk,-}}^{r_{ijk,+}}r^2\sin\theta drd\theta d\phi=\frac{1}{3}(r_{ijk,+}^3-r_{ijk,-}^3)\left(\cos(\theta_{ijk,-})-\cos(\theta_{ijk,+})\right)(\phi_{ijk,+}-\phi_{ijk,-})\,,
\end{equation}
where the integral is carried out between the cell faces, whose coordinates are denoted by $r_{ijk,\pm}$, $\theta_{ijk,\pm}$, and $\phi_{ijk,\pm}$.

The total gas internal energy is given by:
\begin{equation}
U_\mathrm{gas}=\sum_{ijk}\frac{P_{\mathrm{gas},\,ijk}}{\gamma-1}\Delta V_{ijk}\,,
\end{equation}
where $P_\mathrm{gas}$ is the gas pressure and $\gamma=5/3$ is the adiabatic index. The total energy in the radiation is given by:
\begin{equation}
U_\mathrm{rad}=\sum_{ijk}\epsilon_{\mathrm{r},\,ijk}\Delta V_{ijk}\,,
\end{equation}
where $\epsilon_\mathrm{r}$ is the radiation energy density.
The total gravitational binding energy is given by:
\begin{equation}
    E_\mathrm{grav}=\sum_{ijk}-\frac{GM\rho_{ijk}\Delta V_{ijk}}{r_i}\,,
\end{equation}
where we safely ignore the self-gravity of the envelope.

\subsection{Luminosity}
\label{app:luminosity}

We calculate the total bolometric luminosity by integrating the radial radiative flux over the outermost layer of the simulation domain, well above the photosphere. The radial radiative flux in the cell can be calculated from the radiation momentum source term (radial force per unit volume) $f_{\mathrm{rad},\,r}$ in \texttt{Athena++}:
\begin{equation}
    F_{\mathrm{rad},r}=\frac{cf_{\mathrm{rad},\,r}}{\kappa_\mathrm{Rosseland}\rho}\,.
\end{equation}
The total luminosity of the entire star is given by
\begin{equation}
    L_\mathrm{bol}=\frac{4\pi}{\Omega}\sum_{jk}F_{\mathrm{rad},r,N_rjk}\,r^2_{N_rjk}\Delta \Omega_{jk}\,.
\end{equation}
Here, the solid angle for each cell $\Delta\Omega_{jk}$ is given by
\begin{equation}
    \Delta \Omega_{jk}\equiv\int_{\phi_{ijk,-}}^{\phi_{ijk,+}}\int_{\theta_{ijk,-}}^{\theta_{ijk,+}}\sin\theta d\theta d\phi=\left(\cos(\theta_{ijk,-})-\cos(\theta_{ijk,+})\right)(\phi_{ijk,+}-\phi_{ijk,-})\,,
\end{equation}
and the factor $4\pi/\Omega$ scales the wedge luminosity to the whole sphere, where $\Omega=\sum_{jk}\Delta\Omega_{jk}=\pi/3$ is the total solid angle of our simulation domain. 

\subsection{Photosphere and Effective Temperature}
\label{app:photosphere}

As discussed in Section \ref{sec:convection}, the 3D stellar envelope generally has different optical depths at each angular direction. To obtain an angular-averaged photosphere, we calculate the angular-dependent optical depth of each cell by integrating from the outermost cell in the same angular direction:
\begin{equation}
    \tau_{ijk}=\int_{r_i}^{r_{N_r}}\kappa_\mathrm{Rosseland}(r,\theta_j,\phi_k)\rho (r,\theta_j,\phi_k)dr\,.,
\end{equation}
where $\kappa_\mathrm{Rosseland}(r,\theta_j,\phi_k)$ and $\rho (r,\theta_j,\phi_k)$ are sampled using the volume-centered values for each fixed direction $\theta_j$ and $\phi_k$. We then calculate the mean optical depth $\bar{\tau}_i$ at $r_i$ by averaging $\tau_{ijk}$ over angles, weighted by the solid angle $\Delta \Omega_{jk}$:
\begin{equation}
    \bar{\tau}_i = \frac{1}{\Omega}\sum_{jk}\tau_{ijk}\Delta\Omega_{jk}.
\end{equation}

Then, the mean photosphere radius $r_\mathrm{photo,\,3D}$ is calculated by interpolating $r$ over $\bar{\tau}$, such that 
\begin{equation}
    r_\mathrm{photo,\,3D}=r(\bar{\tau}=1)\,.
\end{equation}
The effective temperature is given by:
\begin{equation}
    T_\mathrm{eff}=\left(\frac{L_\mathrm{bol}}{4\pi\sigma_\mathrm{SB} r^2_\mathrm{photo,\,3D}}\right)^{1/4}\,,
\end{equation}
where $\sigma_\mathrm{SB}$ is the Stefan-Boltzmann constant.

\subsection{Radial Profiles}
\label{app:radial_profiles}

The angular-averaged radial profiles are generally calculated by the volume averaged 3D profiles over all angles. For example, the angular-averaged gas density $\bar{\rho}_i$ and temperature $\bar{T}_i$ are given by:
\begin{equation}
    \bar{\rho}_i=\frac{1}{\sum_{jk}\Delta V_{ijk}}\sum_{jk}\rho_{ijk}\Delta V_{ijk}=\frac{1}{\Omega}\sum_{jk}\rho_{ijk}\Delta \Omega_{jk}\,,
\end{equation}
\begin{equation}
    \bar{T}_i=\frac{1}{\sum_{jk}\Delta V_{ijk}}\sum_{jk}T_{ijk}\Delta V_{ijk}=\frac{1}{\Omega}\sum_{jk}T_{ijk}\Delta \Omega_{jk}\,,
\end{equation}
where we used $\Delta V_{ijk}\propto\Delta \Omega_{jk}$ for fixed $i$. The angular-averaged velocity amplitude is calculated using the root mean square of 3D velocities $\mathbf{v}$ in each angular shell:
\begin{equation}
    \bar{v}_i=\sqrt{\frac{1}{\Omega}\sum_{jk}|\mathbf{v}^2_{ijk}|\Delta\Omega_{jk}}\,.
\end{equation}

The total mass flow in the radial direction at $r_i$ is given by:
\begin{equation}
    \dot{M}_{r,\,i}=\frac{4\pi}{\Omega}\sum_{jk}\rho_{ijk}v_{r,\,{ijk}}r_i^2\Delta\Omega_{jk}\,,
\end{equation}
where $v_{r,\,{ijk}}$ is the radial velocity at each cell.

\bibliography{bibliography}{}

\begin{thebibliography}{}
\expandafter\ifx\csname natexlab\endcsname\relax\def\natexlab#1{#1}\fi
\providecommand{\url}[1]{\href{#1}{#1}}
\providecommand{\dodoi}[1]{doi:~\href{http://doi.org/#1}{\nolinkurl{#1}}}
\providecommand{\doeprint}[1]{\href{http://ascl.net/#1}{\nolinkurl{http://ascl.net/#1}}}
\providecommand{\doarXiv}[1]{\href{https://arxiv.org/abs/#1}{\nolinkurl{https://arxiv.org/abs/#1}}}

\bibitem[{C. {Aerts} {et~al.}(2009){Aerts}, {Puls}, {Godart}, \&
  {Dupret}}]{Aerts2009}
{Aerts}, C., {Puls}, J., {Godart}, M., \& {Dupret}, M.-A. 2009,
  \bibinfo{title}{{Collective pulsational velocity broadening due to gravity
  modes as a physical explanation for macroturbulence in hot massive stars},}
  \aap, 508, 409, \dodoi{10.1051/0004-6361/200810471}

\bibitem[{C. {Aerts} {et~al.}(2018){Aerts}, {Bowman}, {S{\'\i}mon-D{\'\i}az},
  {Buysschaert}, {Johnston}, {Moravveji}, {Beck}, {De Cat}, {Triana},
  {Aigrain}, {Castro}, {Huber}, \& {White}}]{Aerts2018}
{Aerts}, C., {Bowman}, D.~M., {S{\'\i}mon-D{\'\i}az}, S., {et~al.} 2018,
  \bibinfo{title}{{K2 photometry and HERMES spectroscopy of the blue supergiant
  {\ensuremath{\rho}} Leo: rotational wind modulation and low-frequency
  waves},} \mnras, 476, 1234, \dodoi{10.1093/mnras/sty308}

\bibitem[{E.~H. {Anders} {et~al.}(2023){Anders}, {Lecoanet}, {Cantiello},
  {Burns}, {Hyatt}, {Kaufman}, {Townsend}, {Brown}, {Vasil}, {Oishi}, \&
  {Jermyn}}]{Anders2023}
{Anders}, E.~H., {Lecoanet}, D., {Cantiello}, M., {et~al.} 2023,
  \bibinfo{title}{{The photometric variability of massive stars due to gravity
  waves excited by core convection},} Nature Astronomy,
  \dodoi{10.1038/s41550-023-02040-7}

\bibitem[{ {Astropy Collaboration} {et~al.}(2013){Astropy Collaboration},
  {Robitaille}, {Tollerud}, {Greenfield}, {Droettboom}, {Bray}, {Aldcroft},
  {Davis}, {Ginsburg}, {Price-Whelan}, {Kerzendorf}, {Conley}, {Crighton},
  {Barbary}, {Muna}, {Ferguson}, {Grollier}, {Parikh}, {Nair}, {Unther},
  {Deil}, {Woillez}, {Conseil}, {Kramer}, {Turner}, {Singer}, {Fox}, {Weaver},
  {Zabalza}, {Edwards}, {Azalee Bostroem}, {Burke}, {Casey}, {Crawford},
  {Dencheva}, {Ely}, {Jenness}, {Labrie}, {Lim}, {Pierfederici}, {Pontzen},
  {Ptak}, {Refsdal}, {Servillat}, \& {Streicher}}]{astropy:2013}
{Astropy Collaboration}, {Robitaille}, T.~P., {Tollerud}, E.~J., {et~al.} 2013,
  \bibinfo{title}{{Astropy: A community Python package for astronomy},} \aap,
  558, A33, \dodoi{10.1051/0004-6361/201322068}

\bibitem[{ {Astropy Collaboration} {et~al.}(2018){Astropy Collaboration},
  {Price-Whelan}, {Sip{\H{o}}cz}, {G{\"u}nther}, {Lim}, {Crawford}, {Conseil},
  {Shupe}, {Craig}, {Dencheva}, {Ginsburg}, {Vand erPlas}, {Bradley},
  {P{\'e}rez-Su{\'a}rez}, {de Val-Borro}, {Aldcroft}, {Cruz}, {Robitaille},
  {Tollerud}, {Ardelean}, {Babej}, {Bach}, {Bachetti}, {Bakanov}, {Bamford},
  {Barentsen}, {Barmby}, {Baumbach}, {Berry}, {Biscani}, {Boquien}, {Bostroem},
  {Bouma}, {Brammer}, {Bray}, {Breytenbach}, {Buddelmeijer}, {Burke},
  {Calderone}, {Cano Rodr{\'\i}guez}, {Cara}, {Cardoso}, {Cheedella}, {Copin},
  {Corrales}, {Crichton}, {D'Avella}, {Deil}, {Depagne}, {Dietrich}, {Donath},
  {Droettboom}, {Earl}, {Erben}, {Fabbro}, {Ferreira}, {Finethy}, {Fox},
  {Garrison}, {Gibbons}, {Goldstein}, {Gommers}, {Greco}, {Greenfield},
  {Groener}, {Grollier}, {Hagen}, {Hirst}, {Homeier}, {Horton}, {Hosseinzadeh},
  {Hu}, {Hunkeler}, {Ivezi{\'c}}, {Jain}, {Jenness}, {Kanarek}, {Kendrew},
  {Kern}, {Kerzendorf}, {Khvalko}, {King}, {Kirkby}, {Kulkarni}, {Kumar},
  {Lee}, {Lenz}, {Littlefair}, {Ma}, {Macleod}, {Mastropietro}, {McCully},
  {Montagnac}, {Morris}, {Mueller}, {Mumford}, {Muna}, {Murphy}, {Nelson},
  {Nguyen}, {Ninan}, {N{\"o}the}, {Ogaz}, {Oh}, {Parejko}, {Parley}, {Pascual},
  {Patil}, {Patil}, {Plunkett}, {Prochaska}, {Rastogi}, {Reddy Janga},
  {Sabater}, {Sakurikar}, {Seifert}, {Sherbert}, {Sherwood-Taylor}, {Shih},
  {Sick}, {Silbiger}, {Singanamalla}, {Singer}, {Sladen}, {Sooley},
  {Sornarajah}, {Streicher}, {Teuben}, {Thomas}, {Tremblay}, {Turner},
  {Terr{\'o}n}, {van Kerkwijk}, {de la Vega}, {Watkins}, {Weaver}, {Whitmore},
  {Woillez}, {Zabalza}, \& {Astropy Contributors}}]{astropy:2018}
{Astropy Collaboration}, {Price-Whelan}, A.~M., {Sip{\H{o}}cz}, B.~M., {et~al.}
  2018, \bibinfo{title}{{The Astropy Project: Building an Open-science Project
  and Status of the v2.0 Core Package},} \aj, 156, 123,
  \dodoi{10.3847/1538-3881/aabc4f}

\bibitem[{ {Astropy Collaboration} {et~al.}(2022){Astropy Collaboration},
  {Price-Whelan}, {Lim}, {Earl}, {Starkman}, {Bradley}, {Shupe}, {Patil},
  {Corrales}, {Brasseur}, {N{"o}the}, {Donath}, {Tollerud}, {Morris},
  {Ginsburg}, {Vaher}, {Weaver}, {Tocknell}, {Jamieson}, {van Kerkwijk},
  {Robitaille}, {Merry}, {Bachetti}, {G{"u}nther}, {Aldcroft},
  {Alvarado-Montes}, {Archibald}, {B{'o}di}, {Bapat}, {Barentsen}, {Baz{'a}n},
  {Biswas}, {Boquien}, {Burke}, {Cara}, {Cara}, {Conroy}, {Conseil}, {Craig},
  {Cross}, {Cruz}, {D'Eugenio}, {Dencheva}, {Devillepoix}, {Dietrich},
  {Eigenbrot}, {Erben}, {Ferreira}, {Foreman-Mackey}, {Fox}, {Freij}, {Garg},
  {Geda}, {Glattly}, {Gondhalekar}, {Gordon}, {Grant}, {Greenfield}, {Groener},
  {Guest}, {Gurovich}, {Handberg}, {Hart}, {Hatfield-Dodds}, {Homeier},
  {Hosseinzadeh}, {Jenness}, {Jones}, {Joseph}, {Kalmbach}, {Karamehmetoglu},
  {Ka{l}uszy{'n}ski}, {Kelley}, {Kern}, {Kerzendorf}, {Koch}, {Kulumani},
  {Lee}, {Ly}, {Ma}, {MacBride}, {Maljaars}, {Muna}, {Murphy}, {Norman},
  {O'Steen}, {Oman}, {Pacifici}, {Pascual}, {Pascual-Granado}, {Patil},
  {Perren}, {Pickering}, {Rastogi}, {Roulston}, {Ryan}, {Rykoff}, {Sabater},
  {Sakurikar}, {Salgado}, {Sanghi}, {Saunders}, {Savchenko}, {Schwardt},
  {Seifert-Eckert}, {Shih}, {Jain}, {Shukla}, {Sick}, {Simpson},
  {Singanamalla}, {Singer}, {Singhal}, {Sinha}, {Sip{H{o}}cz}, {Spitler},
  {Stansby}, {Streicher}, {{{S}}umak}, {Swinbank}, {Taranu}, {Tewary},
  {Tremblay}, {Val-Borro}, {Van Kooten}, {Vasovi{'c}}, {Verma}, {de Miranda
  Cardoso}, {Williams}, {Wilson}, {Winkel}, {Wood-Vasey}, {Xue}, {Yoachim},
  {Zhang}, {Zonca}, \& {Astropy Project Contributors}}]{astropy:2022}
{Astropy Collaboration}, {Price-Whelan}, A.~M., {Lim}, P.~L., {et~al.} 2022,
  \bibinfo{title}{{The Astropy Project: Sustaining and Growing a
  Community-oriented Open-source Project and the Latest Major Release (v5.0) of
  the Core Package},} apj, 935, 167, \dodoi{10.3847/1538-4357/ac7c74}

\bibitem[{J. {Bailey} {et~al.}(2024){Bailey}, {Howarth}, {Cotton},
  {Kedziora-Chudczer}, {De Horta}, {Martell}, {Eldridge}, \&
  {Luckas}}]{Bailey2024}
{Bailey}, J., {Howarth}, I.~D., {Cotton}, D.~V., {et~al.} 2024,
  \bibinfo{title}{{Rapid polarization variations in the O4 supergiant
  {\ensuremath{\zeta}} Puppis},} \mnras, 529, 374,
  \dodoi{10.1093/mnras/stae548}

\bibitem[{E.~P. {Bellinger} {et~al.}(2023){Bellinger}, {de Mink}, {van Rossem},
  \& {Justham}}]{Bellinger2023}
{Bellinger}, E.~P., {de Mink}, S.~E., {van Rossem}, W.~E., \& {Justham}, S.
  2023, \bibinfo{title}{{The Potential of Asteroseismology to Resolve the Blue
  Supergiant Problem},} arXiv e-prints, arXiv:2311.00038,
  \dodoi{10.48550/arXiv.2311.00038}

\bibitem[{D.~M. {Bowman} {et~al.}(2020){Bowman}, {Burssens},
  {Sim{\'o}n-D{\'\i}az}, {Edelmann}, {Rogers}, {Horst}, {R{\"o}pke}, \&
  {Aerts}}]{Bowman2020}
{Bowman}, D.~M., {Burssens}, S., {Sim{\'o}n-D{\'\i}az}, S., {et~al.} 2020,
  \bibinfo{title}{{Photometric detection of internal gravity waves in upper
  main-sequence stars. II. Combined TESS photometry and high-resolution
  spectroscopy},} \aap, 640, A36, \dodoi{10.1051/0004-6361/202038224}

\bibitem[{D.~M. {Bowman} {et~al.}(2019{\natexlab{a}}){Bowman}, {Burssens},
  {Pedersen}, {Johnston}, {Aerts}, {Buysschaert}, {Michielsen}, {Tkachenko},
  {Rogers}, {Edelmann}, {Ratnasingam}, {Sim{\'o}n-D{\'\i}az}, {Castro},
  {Moravveji}, {Pope}, {White}, \& {De Cat}}]{Bowman2019}
{Bowman}, D.~M., {Burssens}, S., {Pedersen}, M.~G., {et~al.}
  2019{\natexlab{a}}, \bibinfo{title}{{Low-frequency gravity waves in blue
  supergiants revealed by high-precision space photometry},} Nature Astronomy,
  3, 760, \dodoi{10.1038/s41550-019-0768-1}

\bibitem[{D.~M. {Bowman} {et~al.}(2019{\natexlab{b}}){Bowman}, {Aerts},
  {Johnston}, {Pedersen}, {Rogers}, {Edelmann}, {Sim{\'o}n-D{\'\i}az}, {Van
  Reeth}, {Buysschaert}, {Tkachenko}, \& {Triana}}]{Bowman2019b}
{Bowman}, D.~M., {Aerts}, C., {Johnston}, C., {et~al.} 2019{\natexlab{b}},
  \bibinfo{title}{{Photometric detection of internal gravity waves in upper
  main-sequence stars. I. Methodology and application to CoRoT targets},} \aap,
  621, A135, \dodoi{10.1051/0004-6361/201833662}

\bibitem[{F. {Bresolin} {et~al.}(2004){Bresolin}, {Pietrzy{\'n}ski}, {Gieren},
  {Kudritzki}, {Przybilla}, \& {Fouqu{\'e}}}]{Bresolin2004}
{Bresolin}, F., {Pietrzy{\'n}ski}, G., {Gieren}, W., {et~al.} 2004,
  \bibinfo{title}{{On the Photometric Variability of Blue Supergiants in NGC
  300 and Its Impact on the Flux-weighted Gravity-Luminosity Relationship},}
  \apj, 600, 182, \dodoi{10.1086/379806}

\bibitem[{A. {\VAN{Burgos}{de}{de} Burgos} {et~al.}(2023){\VAN{Burgos}{de}{de}
  Burgos}, {Sim{\'o}n-D{\'\i}az}, {Urbaneja}, \& {Negueruela}}]{deBurgos2023}
{\VAN{Burgos}{de}{de} Burgos}, A., {Sim{\'o}n-D{\'\i}az}, S., {Urbaneja},
  M.~A., \& {Negueruela}, I. 2023, \bibinfo{title}{{The IACOB project. IX.
  Building a modern empirical database of Galactic O9 - B9 supergiants: Sample
  selection, description, and completeness},} \aap, 674, A212,
  \dodoi{10.1051/0004-6361/202346179}

\bibitem[{M. {Cantiello} {et~al.}(2021){Cantiello}, {Lecoanet}, {Jermyn}, \&
  {Grassitelli}}]{Cantiello2021}
{Cantiello}, M., {Lecoanet}, D., {Jermyn}, A.~S., \& {Grassitelli}, L. 2021,
  \bibinfo{title}{{On the Origin of Stochastic, Low-Frequency Photometric
  Variability in Massive Stars},} \apj, 915, 112,
  \dodoi{10.3847/1538-4357/ac03b0}

\bibitem[{N. {Castro} {et~al.}(2014){Castro}, {Fossati}, {Langer},
  {Sim{\'o}n-D{\'\i}az}, {Schneider}, \& {Izzard}}]{Castro2014}
{Castro}, N., {Fossati}, L., {Langer}, N., {et~al.} 2014, \bibinfo{title}{{The
  spectroscopic Hertzsprung-Russell diagram of Galactic massive stars},} \aap,
  570, L13, \dodoi{10.1051/0004-6361/201425028}

\bibitem[{N. {Castro} {et~al.}(2018){Castro}, {Oey}, {Fossati}, \&
  {Langer}}]{Castro2018}
{Castro}, N., {Oey}, M.~S., {Fossati}, L., \& {Langer}, N. 2018,
  \bibinfo{title}{{The Spectroscopic Hertzsprung-Russell Diagram of Hot Massive
  Stars in the Small Magellanic Cloud},} \apj, 868, 57,
  \dodoi{10.3847/1538-4357/aae6d0}

\bibitem[{D. {Debnath} {et~al.}(2024){Debnath}, {Sundqvist}, {Moens}, {Van der
  Sijpt}, {Verhamme}, \& {Poniatowski}}]{Debnath2024}
{Debnath}, D., {Sundqvist}, J.~O., {Moens}, N., {et~al.} 2024,
  \bibinfo{title}{{2D unified atmosphere and wind simulations of O-type
  stars},} \aap, 684, A177, \dodoi{10.1051/0004-6361/202348206}

\bibitem[{J.~A. {Goldberg} {et~al.}(2022){Goldberg}, {Jiang}, \&
  {Bildsten}}]{Goldberg2022}
{Goldberg}, J.~A., {Jiang}, Y.-F., \& {Bildsten}, L. 2022,
  \bibinfo{title}{{Numerical Simulations of Convective Three-dimensional Red
  Supergiant Envelopes},} \apj, 929, 156, \dodoi{10.3847/1538-4357/ac5ab3}

\bibitem[{L. {Grassitelli} {et~al.}(2015){Grassitelli}, {Fossati},
  {Sim{\'o}n-Di{\'a}z}, {Langer}, {Castro}, \& {Sanyal}}]{Grassitelli2015}
{Grassitelli}, L., {Fossati}, L., {Sim{\'o}n-Di{\'a}z}, S., {et~al.} 2015,
  \bibinfo{title}{{Observational Consequences of Turbulent Pressure in the
  Envelopes of Massive Stars},} \apjl, 808, L31,
  \dodoi{10.1088/2041-8205/808/1/L31}

\bibitem[{N. {Grevesse} \& A.~J. {Sauval}(1998){Grevesse} \&
  {Sauval}}]{Grevesse1998}
{Grevesse}, N., \& {Sauval}, A.~J. 1998, \bibinfo{title}{{Standard Solar
  Composition},} \ssr, 85, 161, \dodoi{10.1023/A:1005161325181}

\bibitem[{C.~R. Harris {et~al.}(2020)Harris, Millman, van~der Walt, Gommers,
  Virtanen, Cournapeau, Wieser, Taylor, Berg, Smith, Kern, Picus, Hoyer, van
  Kerkwijk, Brett, Haldane, del R{\'{i}}o, Wiebe, Peterson,
  G{\'{e}}rard-Marchant, Sheppard, Reddy, Weckesser, Abbasi, Gohlke, \&
  Oliphant}]{Harris2020}
Harris, C.~R., Millman, K.~J., van~der Walt, S.~J., {et~al.} 2020,
  \bibinfo{title}{Array programming with {NumPy},} Nature, 585, 357,
  \dodoi{10.1038/s41586-020-2649-2}

\bibitem[{I.~D. {Howarth} {et~al.}(1997){Howarth}, {Siebert}, {Hussain}, \&
  {Prinja}}]{Howarth1997}
{Howarth}, I.~D., {Siebert}, K.~W., {Hussain}, G. A.~J., \& {Prinja}, R.~K.
  1997, \bibinfo{title}{{Cross-correlation characteristics of OB stars from IUE
  spectroscopy},} \mnras, 284, 265, \dodoi{10.1093/mnras/284.2.265}

\bibitem[{C.~A. {Iglesias} \& F.~J. {Rogers}(1993){Iglesias} \&
  {Rogers}}]{Iglesias1993}
{Iglesias}, C.~A., \& {Rogers}, F.~J. 1993, \bibinfo{title}{{Radiative
  Opacities for Carbon- and Oxygen-rich Mixtures},} \apj, 412, 752,
  \dodoi{10.1086/172958}

\bibitem[{C.~A. {Iglesias} \& F.~J. {Rogers}(1996){Iglesias} \&
  {Rogers}}]{Iglesias1996}
{Iglesias}, C.~A., \& {Rogers}, F.~J. 1996, \bibinfo{title}{{Updated Opal
  Opacities},} \apj, 464, 943, \dodoi{10.1086/177381}

\bibitem[{J.~M. {Jenkins} {et~al.}(2016){Jenkins}, {Twicken}, {McCauliff},
  {Campbell}, {Sanderfer}, {Lung}, {Mansouri-Samani}, {Girouard}, {Tenenbaum},
  {Klaus}, {Smith}, {Caldwell}, {Chacon}, {Henze}, {Heiges}, {Latham},
  {Morgan}, {Swade}, {Rinehart}, \& {Vanderspek}}]{Jenkins2016}
{Jenkins}, J.~M., {Twicken}, J.~D., {McCauliff}, S., {et~al.} 2016,
  \bibinfo{title}{{The TESS science processing operations center},} in Society
  of Photo-Optical Instrumentation Engineers (SPIE) Conference Series, Vol.
  9913, Software and Cyberinfrastructure for Astronomy IV, ed. G.~{Chiozzi} \&
  J.~C. {Guzman}, 99133E, \dodoi{10.1117/12.2233418}

\bibitem[{A.~S. {Jermyn} {et~al.}(2023){Jermyn}, {Bauer}, {Schwab}, {Farmer},
  {Ball}, {Bellinger}, {Dotter}, {Joyce}, {Marchant}, {Mombarg}, {Wolf}, {Sunny
  Wong}, {Cinquegrana}, {Farrell}, {Smolec}, {Thoul}, {Cantiello}, {Herwig},
  {Toloza}, {Bildsten}, {Townsend}, \& {Timmes}}]{Jermyn2023}
{Jermyn}, A.~S., {Bauer}, E.~B., {Schwab}, J., {et~al.} 2023,
  \bibinfo{title}{{Modules for Experiments in Stellar Astrophysics (MESA):
  Time-dependent Convection, Energy Conservation, Automatic Differentiation,
  and Infrastructure},} \apjs, 265, 15, \dodoi{10.3847/1538-4365/acae8d}

\bibitem[{Y.-F. {Jiang}(2021){Jiang}}]{Jiang2021}
{Jiang}, Y.-F. 2021, \bibinfo{title}{{An Implicit Finite Volume Scheme to Solve
  the Time-dependent Radiation Transport Equation Based on Discrete
  Ordinates},} \apjs, 253, 49, \dodoi{10.3847/1538-4365/abe303}

\bibitem[{A. {Kaufer} {et~al.}(1997){Kaufer}, {Stahl}, {Wolf}, {Fullerton},
  {Gaeng}, {Gummersbach}, {Jankovics}, {Kovacs}, {Mandel}, {Peitz}, {Rivinius},
  \& {Szeifert}}]{Kaufer1997}
{Kaufer}, A., {Stahl}, O., {Wolf}, B., {et~al.} 1997,
  \bibinfo{title}{{Long-term spectroscopic monitoring of BA-type supergiants.
  III. Variability of photospheric lines.},} \aap, 320, 273

\bibitem[{M. {Kourniotis} {et~al.}(2025){Kourniotis}, {Cidale}, {Kraus}, {Ruiz
  Diaz}, \& {Alberici Adam}}]{Kourniotis2025}
{Kourniotis}, M., {Cidale}, L.~S., {Kraus}, M., {Ruiz Diaz}, M.~A., \&
  {Alberici Adam}, A. 2025, \bibinfo{title}{{Variability of Galactic blue
  supergiants observed with TESS},} \aap, 697, A152,
  \dodoi{10.1051/0004-6361/202452360}

\bibitem[{J. {Krti{\v{c}}ka} \& A. {Feldmeier}(2018){Krti{\v{c}}ka} \&
  {Feldmeier}}]{Krticka2018}
{Krti{\v{c}}ka}, J., \& {Feldmeier}, A. 2018, \bibinfo{title}{{Light variations
  due to the line-driven wind instability and wind blanketing in O stars},}
  \aap, 617, A121, \dodoi{10.1051/0004-6361/201731614}

\bibitem[{J. {Krti{\v{c}}ka} \& A. {Feldmeier}(2021){Krti{\v{c}}ka} \&
  {Feldmeier}}]{Krticka2021}
{Krti{\v{c}}ka}, J., \& {Feldmeier}, A. 2021, \bibinfo{title}{{Stochastic light
  variations in hot stars from wind instability: finding photometric signatures
  and testing against the TESS data},} \aap, 648, A79,
  \dodoi{10.1051/0004-6361/202040148}

\bibitem[{ {Lightkurve Collaboration} {et~al.}(2018){Lightkurve Collaboration},
  {Cardoso}, {Hedges}, {Gully-Santiago}, {Saunders}, {Cody}, {Barclay}, {Hall},
  {Sagear}, {Turtelboom}, {Zhang}, {Tzanidakis}, {Mighell}, {Coughlin}, {Bell},
  {Berta-Thompson}, {Williams}, {Dotson}, \& {Barentsen}}]{Lightkurve}
{Lightkurve Collaboration}, {Cardoso}, J.~V.~d.~M., {Hedges}, C., {et~al.}
  2018, {Lightkurve: Kepler and TESS time series analysis in Python},,
  Astrophysics Source Code Library \doeprint{1812.013}

\bibitem[{L.~B. {Lucy}(1976){Lucy}}]{Lucy1976}
{Lucy}, L.~B. 1976, \bibinfo{title}{{An analysis of the variable radial
  velocity of Alpha Cygni.},} \apj, 206, 499, \dodoi{10.1086/154405}

\bibitem[{J.-Z. {Ma} {et~al.}(2025){Ma}, {Justham}, {Pakmor}, {Chiavassa},
  {Ryu}, \& {de Mink}}]{Ma2025}
{Ma}, J.-Z., {Justham}, S., {Pakmor}, R., {et~al.} 2025,
  \bibinfo{title}{{AREPO-RSG: Aspherical Circumstellar Material and Winds from
  Pulsating Dusty Red Supergiants in Global 3D Radiation Hydrodynamic
  Simulations},} arXiv e-prints, arXiv:2510.14875,
  \dodoi{10.48550/arXiv.2510.14875}

\bibitem[{L. {Ma} {et~al.}(2024){Ma}, {Johnston}, {Bellinger}, \& {de
  Mink}}]{Ma2024}
{Ma}, L., {Johnston}, C., {Bellinger}, E.~P., \& {de Mink}, S.~E. 2024,
  \bibinfo{title}{{Variability of Blue Supergiants in the LMC with TESS},}
  \apj, 966, 196, \dodoi{10.3847/1538-4357/ad38bc}

\bibitem[{A. {Menon} {et~al.}(2024){Menon}, {Ercolino}, {Urbaneja}, {Lennon},
  {Herrero}, {Hirai}, {Langer}, {Schootemeijer}, {Chatzopoulos}, {Frank}, \&
  {Shiber}}]{Menon2024}
{Menon}, A., {Ercolino}, A., {Urbaneja}, M.~A., {et~al.} 2024,
  \bibinfo{title}{{Evidence for Evolved Stellar Binary Mergers in Observed
  B-type Blue Supergiants},} \apjl, 963, L42, \dodoi{10.3847/2041-8213/ad2074}

\bibitem[{E. {Moravveji} {et~al.}(2012){Moravveji}, {Guinan}, {Shultz},
  {Williamson}, \& {Moya}}]{Moravveji2012}
{Moravveji}, E., {Guinan}, E.~F., {Shultz}, M., {Williamson}, M.~H., \& {Moya},
  A. 2012, \bibinfo{title}{{Asteroseismology of the nearby SN-II Progenitor:
  Rigel. I. The MOST High-precision Photometry and Radial Velocity
  Monitoring},} \apj, 747, 108, \dodoi{10.1088/0004-637X/747/2/108}

\bibitem[{B. {Paxton} {et~al.}(2011){Paxton}, {Bildsten}, {Dotter}, {Herwig},
  {Lesaffre}, \& {Timmes}}]{Paxton2011}
{Paxton}, B., {Bildsten}, L., {Dotter}, A., {et~al.} 2011,
  \bibinfo{title}{{Modules for Experiments in Stellar Astrophysics (MESA)},}
  \apjs, 192, 3, \dodoi{10.1088/0067-0049/192/1/3}

\bibitem[{B. {Paxton} {et~al.}(2013){Paxton}, {Cantiello}, {Arras}, {Bildsten},
  {Brown}, {Dotter}, {Mankovich}, {Montgomery}, {Stello}, {Timmes}, \&
  {Townsend}}]{Paxton2013}
{Paxton}, B., {Cantiello}, M., {Arras}, P., {et~al.} 2013,
  \bibinfo{title}{{Modules for Experiments in Stellar Astrophysics (MESA):
  Planets, Oscillations, Rotation, and Massive Stars},} \apjs, 208, 4,
  \dodoi{10.1088/0067-0049/208/1/4}

\bibitem[{B. {Paxton} {et~al.}(2015){Paxton}, {Marchant}, {Schwab}, {Bauer},
  {Bildsten}, {Cantiello}, {Dessart}, {Farmer}, {Hu}, {Langer}, {Townsend},
  {Townsley}, \& {Timmes}}]{Paxton2015}
{Paxton}, B., {Marchant}, P., {Schwab}, J., {et~al.} 2015,
  \bibinfo{title}{{Modules for Experiments in Stellar Astrophysics (MESA):
  Binaries, Pulsations, and Explosions},} \apjs, 220, 15,
  \dodoi{10.1088/0067-0049/220/1/15}

\bibitem[{B. {Paxton} {et~al.}(2018){Paxton}, {Schwab}, {Bauer}, {Bildsten},
  {Blinnikov}, {Duffell}, {Farmer}, {Goldberg}, {Marchant}, {Sorokina},
  {Thoul}, {Townsend}, \& {Timmes}}]{Paxton2018}
{Paxton}, B., {Schwab}, J., {Bauer}, E.~B., {et~al.} 2018,
  \bibinfo{title}{{Modules for Experiments in Stellar Astrophysics (MESA):
  Convective Boundaries, Element Diffusion, and Massive Star Explosions},}
  \apjs, 234, 34, \dodoi{10.3847/1538-4365/aaa5a8}

\bibitem[{B. {Paxton} {et~al.}(2019){Paxton}, {Smolec}, {Schwab}, {Gautschy},
  {Bildsten}, {Cantiello}, {Dotter}, {Farmer}, {Goldberg}, {Jermyn}, {Kanbur},
  {Marchant}, {Thoul}, {Townsend}, {Wolf}, {Zhang}, \& {Timmes}}]{Paxton2019}
{Paxton}, B., {Smolec}, R., {Schwab}, J., {et~al.} 2019,
  \bibinfo{title}{{Modules for Experiments in Stellar Astrophysics (MESA):
  Pulsating Variable Stars, Rotation, Convective Boundaries, and Energy
  Conservation},} \apjs, 243, 10, \dodoi{10.3847/1538-4365/ab2241}

\bibitem[{M.~G. {Pedersen} \& L. {Bildsten}(2025){Pedersen} \&
  {Bildsten}}]{Pedersen2025}
{Pedersen}, M.~G., \& {Bildsten}, L. 2025, \bibinfo{title}{{Stochastic
  low-frequency variability of 50 massive stars in the Cygnus OB associations
  and the Small Magellanic Cloud},} \mnras, 539, 2742,
  \dodoi{10.1093/mnras/staf661}

\bibitem[{M.~G. {Pedersen} {et~al.}(2019){Pedersen}, {Chowdhury}, {Johnston},
  {Bowman}, {Aerts}, {Handler}, {De Cat}, {Neiner}, {David-Uraz}, {Buzasi},
  {Tkachenko}, {Sim{\'o}n-D{\'\i}az}, {Moravveji}, {Sikora}, {Mirouh},
  {Lovekin}, {Cantiello}, {Daszy{\'n}ska-Daszkiewicz}, {Pigulski},
  {Vanderspek}, \& {Ricker}}]{Pedersen2019}
{Pedersen}, M.~G., {Chowdhury}, S., {Johnston}, C., {et~al.} 2019,
  \bibinfo{title}{{Diverse Variability of O and B Stars Revealed from 2-minute
  Cadence Light Curves in Sectors 1 and 2 of the TESS Mission: Selection of an
  Asteroseismic Sample},} \apjl, 872, L9, \dodoi{10.3847/2041-8213/ab01e1}

\bibitem[{T. {Ramiaramanantsoa} {et~al.}(2018){Ramiaramanantsoa}, {Moffat},
  {Harmon}, {Ignace}, {St-Louis}, {Vanbeveren}, {Shenar}, {Pablo},
  {Richardson}, {Howarth}, {Stevens}, {Piaulet}, {St-Jean}, {Eversberg},
  {Pigulski}, {Popowicz}, {Kuschnig}, {Zoc{\l}o{\'n}ska}, {Buysschaert},
  {Handler}, {Weiss}, {Wade}, {Rucinski}, {Zwintz}, {Luckas}, {Heathcote},
  {Cacella}, {Powles}, {Locke}, {Bohlsen}, {Chen{\'e}}, {Miszalski}, {Waldron},
  {Kotze}, {Kotze}, \& {B{\"o}hm}}]{Ramiaramanantsoa2018}
{Ramiaramanantsoa}, T., {Moffat}, A. F.~J., {Harmon}, R., {et~al.} 2018,
  \bibinfo{title}{{BRITE-Constellation high-precision time-dependent photometry
  of the early O-type supergiant {\ensuremath{\zeta}} Puppis unveils the
  photospheric drivers of its small- and large-scale wind structures},} \mnras,
  473, 5532, \dodoi{10.1093/mnras/stx2671}

\bibitem[{G.~R. {Ricker} {et~al.}(2015){Ricker}, {Winn}, {Vanderspek},
  {Latham}, {Bakos}, {Bean}, {Berta-Thompson}, {Brown}, {Buchhave}, {Butler},
  {Butler}, {Chaplin}, {Charbonneau}, {Christensen-Dalsgaard}, {Clampin},
  {Deming}, {Doty}, {De Lee}, {Dressing}, {Dunham}, {Endl}, {Fressin}, {Ge},
  {Henning}, {Holman}, {Howard}, {Ida}, {Jenkins}, {Jernigan}, {Johnson},
  {Kaltenegger}, {Kawai}, {Kjeldsen}, {Laughlin}, {Levine}, {Lin}, {Lissauer},
  {MacQueen}, {Marcy}, {McCullough}, {Morton}, {Narita}, {Paegert}, {Palle},
  {Pepe}, {Pepper}, {Quirrenbach}, {Rinehart}, {Sasselov}, {Sato}, {Seager},
  {Sozzetti}, {Stassun}, {Sullivan}, {Szentgyorgyi}, {Torres}, {Udry}, \&
  {Villasenor}}]{TESS}
{Ricker}, G.~R., {Winn}, J.~N., {Vanderspek}, R., {et~al.} 2015,
  \bibinfo{title}{{Transiting Exoplanet Survey Satellite (TESS)},} Journal of
  Astronomical Telescopes, Instruments, and Systems, 1, 014003,
  \dodoi{10.1117/1.JATIS.1.1.014003}

\bibitem[{T.~M. {Rogers} {et~al.}(2013){Rogers}, {Lin}, {McElwaine}, \&
  {Lau}}]{Rogers2013}
{Rogers}, T.~M., {Lin}, D.~N.~C., {McElwaine}, J.~N., \& {Lau}, H.~H.~B. 2013,
  \bibinfo{title}{{Internal Gravity Waves in Massive Stars: Angular Momentum
  Transport},} \apj, 772, 21, \dodoi{10.1088/0004-637X/772/1/21}

\bibitem[{R.~S.~I. {Ryans} {et~al.}(2002){Ryans}, {Dufton}, {Rolleston},
  {Lennon}, {Keenan}, {Smoker}, \& {Lambert}}]{Ryans2002}
{Ryans}, R.~S.~I., {Dufton}, P.~L., {Rolleston}, W.~R.~J., {et~al.} 2002,
  \bibinfo{title}{{Macroturbulent and rotational broadening in the spectra of
  B-type supergiants},} \mnras, 336, 577,
  \dodoi{10.1046/j.1365-8711.2002.05780.x}

\bibitem[{F.~R.~N. {Schneider} {et~al.}(2019){Schneider}, {Ohlmann},
  {Podsiadlowski}, {R{\"o}pke}, {Balbus}, {Pakmor}, \&
  {Springel}}]{Schneider2019}
{Schneider}, F. R.~N., {Ohlmann}, S.~T., {Podsiadlowski}, P., {et~al.} 2019,
  \bibinfo{title}{{Stellar mergers as the origin of magnetic massive stars},}
  \nat, 574, 211, \dodoi{10.1038/s41586-019-1621-5}

\bibitem[{W.~C. {Schultz} {et~al.}(2022){Schultz}, {Bildsten}, \&
  {Jiang}}]{Schultz2022}
{Schultz}, W.~C., {Bildsten}, L., \& {Jiang}, Y.-F. 2022,
  \bibinfo{title}{{Stochastic Low-frequency Variability in Three-dimensional
  Radiation Hydrodynamical Models of Massive Star Envelopes},} \apjl, 924, L11,
  \dodoi{10.3847/2041-8213/ac441f}

\bibitem[{W.~C. {Schultz} {et~al.}(2023){Schultz}, {Bildsten}, \&
  {Jiang}}]{Schultz2023a}
{Schultz}, W.~C., {Bildsten}, L., \& {Jiang}, Y.-F. 2023,
  \bibinfo{title}{{Turbulence-supported Massive Star Envelopes},} \apjl, 951,
  L42, \dodoi{10.3847/2041-8213/acdf50}

\bibitem[{N. {Serebriakova} {et~al.}(2024){Serebriakova}, {Tkachenko}, \&
  {Aerts}}]{Serebriakova2024}
{Serebriakova}, N., {Tkachenko}, A., \& {Aerts}, C. 2024, \bibinfo{title}{{The
  ESO UVES/FEROS Large Programs of TESS OB pulsators: II. The physical origin
  of macroturbulence},} \aap, 692, A245, \dodoi{10.1051/0004-6361/202451581}

\bibitem[{N. {Serebriakova} {et~al.}(2023){Serebriakova}, {Tkachenko},
  {Gebruers}, {Bowman}, {Van Reeth}, {Mahy}, {Burssens}, {IJspeert}, {Sana}, \&
  {Aerts}}]{Serebriakova2023}
{Serebriakova}, N., {Tkachenko}, A., {Gebruers}, S., {et~al.} 2023,
  \bibinfo{title}{{The ESO UVES/FEROS Large Programs of TESS OB pulsators. I.
  Global stellar parameters from high-resolution spectroscopy},} \aap, 676,
  A85, \dodoi{10.1051/0004-6361/202346108}

\bibitem[{J.~H. {Shiode} {et~al.}(2013){Shiode}, {Quataert}, {Cantiello}, \&
  {Bildsten}}]{Shiode2013}
{Shiode}, J.~H., {Quataert}, E., {Cantiello}, M., \& {Bildsten}, L. 2013,
  \bibinfo{title}{{The observational signatures of convectively excited gravity
  modes in main-sequence stars},} \mnras, 430, 1736,
  \dodoi{10.1093/mnras/sts719}

\bibitem[{S. {Sim{\'o}n-D{\'\i}az} {et~al.}(2010){Sim{\'o}n-D{\'\i}az},
  {Herrero}, {Uytterhoeven}, {Castro}, {Aerts}, \& {Puls}}]{SimonDiaz2010}
{Sim{\'o}n-D{\'\i}az}, S., {Herrero}, A., {Uytterhoeven}, K., {et~al.} 2010,
  \bibinfo{title}{{Observational Evidence for a Correlation Between
  Macroturbulent Broadening and Line-profile Variations in OB Supergiants},}
  \apjl, 720, L174, \dodoi{10.1088/2041-8205/720/2/L174}

\bibitem[{J.~M. {Stone} {et~al.}(2020){Stone}, {Tomida}, {White}, \&
  {Felker}}]{Stone2020}
{Stone}, J.~M., {Tomida}, K., {White}, C.~J., \& {Felker}, K.~G. 2020,
  \bibinfo{title}{{The Athena++ Adaptive Mesh Refinement Framework: Design and
  Magnetohydrodynamic Solvers},} \apjs, 249, 4,
  \dodoi{10.3847/1538-4365/ab929b}

\bibitem[{P. Virtanen {et~al.}(2020)Virtanen, Gommers, Oliphant, Haberland,
  Reddy, Cournapeau, Burovski, Peterson, Weckesser, Bright, {van der Walt},
  Brett, Wilson, Millman, Mayorov, Nelson, Jones, Kern, Larson, Carey, Polat,
  Feng, Moore, {VanderPlas}, Laxalde, Perktold, Cimrman, Henriksen, Quintero,
  Harris, Archibald, Ribeiro, Pedregosa, {van Mulbregt}, \& {SciPy 1.0
  Contributors}}]{scipy}
Virtanen, P., Gommers, R., Oliphant, T.~E., {et~al.} 2020,
  \bibinfo{title}{{{SciPy} 1.0: Fundamental Algorithms for Scientific Computing
  in Python},} Nature Methods, 17, 261, \dodoi{10.1038/s41592-019-0686-2}

\end{thebibliography}
\bibliographystyle{aasjournalv7}

\end{CJK*}
\end{document}